\documentclass[reprint,prc,onecolumn,preprintnumbers,superscriptaddress,longbibliography]{revtex4-2}
\usepackage{graphicx} % Required for inserting images
\usepackage{orcidlink}
\usepackage{graphicx}% Include figure files
\usepackage{dcolumn}% Align table columns on decimal point
\usepackage{bm}% bold math
\usepackage{textgreek}
\usepackage{placeins}
\usepackage{braket}
\usepackage{physics}
\usepackage{overpic}
\usepackage{amssymb}
\usepackage{amsmath}
\usepackage{subfigure}
\usepackage{qcircuit}
\usepackage{tikz}
\usepackage{adjustbox}
\usepackage{longtable}
\usepackage{soul}
\usepackage{tensor}
\usepackage{youngtab}
\usepackage{cleveref}
\newcommand{\wh}{{\ydiagram[]{1}}}
\newcommand{\bl}{{\ydiagram[*(black)]{1}}}
\newcommand{\whwhs}{{\ydiagram[]{2}}}
\newcommand{\blbls}{{\ydiagram[*(black)]{2}}}
\newcommand{\whwha}{{\ydiagram[]{1,1}}}
\newcommand{\blbla}{{\ydiagram[*(black)]{1,1}}}
\newcommand{\blwh}{{\ydiagram[*(white)]{1+1} * [*(black)]{1}}}

\usepackage[smalltableaux]{ytableau}

\usepackage{tikz}

\begin{document}

\title{Efficient Truncations of SU($N_c$) Lattice Gauge Theory for Quantum Simulation}
%\title{SU($N_c$) Lattice Gauge Theory Hamiltonians to Order $1/N_c$}
\author{Anthony N. Ciavarella \,\orcidlink{0000-0003-3918-4110}}
\email{anciavarella@lbl.gov}
\affiliation{Physics Division, Lawrence Berkeley National Laboratory, Berkeley, California 94720, USA}
\author{Ivan M. Burbano \,\orcidlink{0000-0002-3792-1773}}
\email{ivan\_burbano@berkeley.edu}
\affiliation{Physics Division, Lawrence Berkeley National Laboratory, Berkeley, California 94720, USA}
\affiliation{Department of Physics, University of California, Berkeley, Berkeley, CA 94720, USA}
\affiliation{Nuclear Science Division, Lawrence Berkeley National Laboratory, Berkeley, CA 94720, USA}
\author{Christian W. Bauer \,\orcidlink{0000-0001-9820-5810}}
\email{cwbauer@lbl.gov}
\affiliation{Physics Division, Lawrence Berkeley National Laboratory, Berkeley, California 94720, USA}
\affiliation{Department of Physics, University of California, Berkeley, Berkeley, CA 94720, USA}

\date{\today}

\begin{abstract}
    Quantum simulations of lattice gauge theories offer the potential to directly study the non-perturbative dynamics of quantum chromodynamics, but naive analyses suggest that they require large computational resources. Large \texorpdfstring{$N_c$}{Nc} expansions are performed to order \texorpdfstring{$1/N_c$}{1/Nc} to simplify the Hamiltonian of pure SU(\texorpdfstring{$N_c$}{Nc}) lattice gauge theories. A reformulation of the electric basis is introduced with a truncation strategy based on the construction of local Krylov subspaces with plaquette operators. Numerical simulations show that these truncated Hamiltonians are consistent with traditional lattice calculations at relatively small couplings. It is shown that the computational resources required for quantum simulation of time evolution generated by these Hamiltonians is \texorpdfstring{$17-19$}{17-19} orders of magnitude smaller than previous approaches, provided that the truncations in this work can reach lattice spacings in 3D comparable to the 2D simulations performed.
\end{abstract}

\maketitle

\tableofcontents

\section{Introduction}
The simulation of lattice gauge theories on quantum computers offers the potential to directly probe the real time dynamics of strongly coupled gauge theories~\cite{PhysRevD.10.2445,Feynman:1981tf}. Applying this to quantum chromodynamics (QCD) will enable direct simulations of the process of hadronization~\cite{Bauer:2022hpo,Humble:2022vtm,Beck:2023xhh,Bauer:2021gup}. Simulations of real time dynamics will also enable computations of observables relevant to understanding the quark-gluon plasma in the early universe such as the shear viscosity~\cite{Cohen:2021imf,Turro:2024pxu}. Inelastic scattering amplitudes can also be computed for processes outside the reach of Luscher's method~\cite{Luscher:1985dn,Luscher:1986pf,Ciavarella:2020vqm,Briceno:2020rar,Briceno:2021aiw,Carrillo:2024chu}. The impact of quantum simulation on high energy physics is not restricted to the dynamics of QCD. Other SU($N_c$) gauge theories have been proposed as models for grand unification~\cite{Georgi:1974sy,Masiero:1982fe,Baez:2009dj} or as theories of dark matter~\cite{Hardy:2014mqa,Forestell:2016qhc,Francis:2018xjd,Carenza:2022pjd,McKeen:2024trt}. Quantum simulations of the non-perturbative aspects of these theories can aid in searches for new physics beyond the standard model. 

The potential of quantum simulation to study real-time dynamics has motivated new developments in Hamiltonian lattice gauge theory~\cite{Kogut:1974ag,Kogut:1979wt,Jones:1979av,Kogut:1980za,Chin:1986fe,Chin:1987jt,Chin:1987at,Long:1988qe}. Direct attempts to encode lattice gauge theories onto quantum computers have impractically large resource requirements~\cite{Byrnes:2005qx,Kan:2021xfc,Rhodes:2024zbr}. More efficient encodings of the electric basis of gauge theories can be found by exploiting Gauss's law to reduce the number of degrees of freedom that need to be mapped onto a quantum processor~\cite{Banuls:2017ena,Klco:2019evd,Zohar:2019ygc,Ciavarella:2021nmj,Lewis:2019wfx,Raychowdhury:2018osk,Raychowdhury:2019iki,Stryker:2020vls,Kadam:2022ipf,Kadam:2023gpn,Kadam:2024zkj,Zache:2023dko,Kan:2021nyu,Ciavarella:2022zhe,Ciavarella:2024fzw,Muller:2023nnk,Kavaki:2024ijd,Hayata:2023bgh,Rigobello:2023ype,Fontana:2024rux,Balaji:2025afl,Illa:2025dou}. Alternatively, one could use discrete subgroups to encode gauge fields~\cite{Lamm:2019bik,Alexandru:2019nsa,Alam:2021uuq,Ji:2022qvr,Gustafson:2022xdt,Gustafson:2023swx,Gustafson:2023kvd,Gustafson:2024kym,Assi:2024pdn,Hartung:2022hoz}. Hybrid approaches that mix these two approaches have also been developed~\cite{Bauer:2021gek,DAndrea:2023qnr,Jakobs:2023lpp,Garofalo:2023zkd,Grabowska:2024emw,Burbano:2024uvn,Jakobs:2025rvz}. Other approaches using orbifold Hamiltonians~\cite{Buser:2020cvn,Bergner:2024qjl,Halimeh:2024bth} and quantum link models have also been explored~\cite{Brower:1997ha,Brower:2003vy,Liu:2023lsr,Chandrasekharan:2025pil,Chandrasekharan:2025smw,Alexandru:2023qzd}.

These developments have enabled simulations of low dimensional lattice gauge theories on existing quantum computers. The dynamics of the Schwinger model has been simulated with lattice sizes reaching $112$ sites~\cite{Martinez:2016yna,Klco:2018kyo,Yang:2020yer,Zhou:2021kdl,Nguyen:2021hyk,Su:2022glk,Zhang:2023hzr,Angelides:2023noe,Mildenberger:2022jqr,Charles:2023zbl,Mueller:2022xbg,De:2024smi,Davoudi:2024wyv,Guo:2024tnb,Farrell:2023fgd,Farrell:2024fit}. Other $1+1D$ non-Abelian theories have been simulated as well~\cite{Atas:2021ext,Atas:2022dqm,Than:2024zaj,Farrell:2022wyt,Farrell:2022vyh,Ciavarella:2023mfc,Ciavarella:2024lsp}. The quantum simulation of higher dimensional gauge theories is complicated by the presence of plaquette operators. Limited simulations have been performed on two-dimensional lattices, restricted mostly to smaller sizes~\cite{Klco:2019evd,Paulson:2020zjd,Ciavarella:2021nmj,Ciavarella:2021lel,ARahman:2022tkr,Mendicelli:2022ntz,Turro:2024pxu,Li:2024lrl,Halimeh:2024ref,Gupta:2024gnw,Crippa:2024hso,Kavaki:2025hcu}. The first quantum simulation of an SU($3$) lattice gauge theory on a non-trivial two dimensional lattice was performed by simplifying the Hamiltonian with the use of a large $N_c$ expansion~\cite{Ciavarella:2024fzw}.

The large $N_c$ expansion of gauge theories has a long history of providing insight into gauge theories~\cite{Witten:1979pi,Manohar:1998xv,Lucini:2012gg}. This expansion correctly captures qualitative features of QCD such as the OZI rule and the emergent Wigner SU($4$) symmetry of nuclear interactions~\cite{tHooft:1973alw,Kaplan:1995yg}. It is also an essential ingredient in event generators of collider physics~\cite{Sjostrand:2006za,Bahr:2008pv}. Applications of the large $N_c$ expansion to lattice gauge theory actions have shown that in various parameter regimes, the entire lattice can be described by a single plaquette~\cite{Eguchi:1982nm,Bhanot:1982sh,Bringoltz:2008av,Gonzalez-Arroyo:2010omx}. Recent work has demonstrated how restricting to the leading order large $N_c$ piece of the Kogut-Susskind Hamiltonian can dramatically simplify the implementation of quantum simulation of lattice gauge theories~\cite{Ciavarella:2024fzw}. 

In this work, it is shown how sub-leading terms in the large $N_c$ expansion of the Kogut-Susskind Hamiltonian can be included in quantum simulations based on the large $N_c$ limit. Truncations of the electric basis based on constructing local Krylov subspaces are introduced. Numerical simulations of SU($3$) lattice gauge theory in $2+1D$ are performed to develop an understanding of what lattice spacings can be reached with this approach. The resources needed to perform a physically relevant quantum simulation with these truncations are estimated and found to be several orders of magnitude smaller than estimates based on direct encodings of lattice gauge theories.

\section{Formulation of SU\texorpdfstring{$(N_c)$}{(Nc)} Lattice Gauge Theory Hamiltonians in large \texorpdfstring{$N_c$}{(Nc)} expansion to order \texorpdfstring{$1/N_c$}{(1/Nc)}}

\subsection{Electric basis choices}
\label{sec:electric_basis}

\begin{figure}
    \centering
    \includegraphics[width=0.3\linewidth]{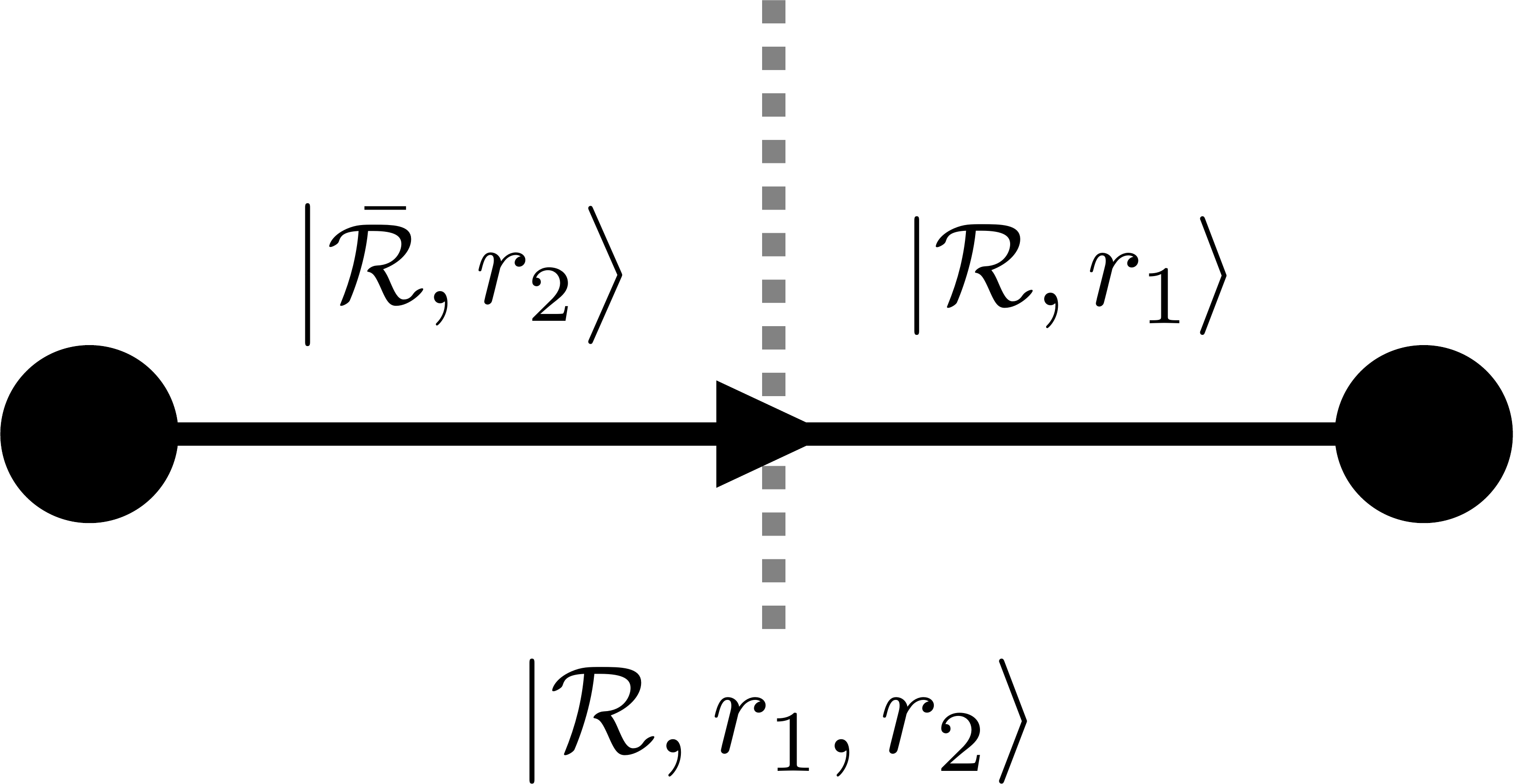}
    \caption{A generic link is shown equipped with a state $\ket{\mathcal{R},r_1,r_2}$. The target half-link is equipped with the vector $\ket{\mathcal{R},r_1}$, while the source half-link is equipped with the vector $\ket{\bar{\mathcal{R}},r_2}\equiv\bra{\mathcal{R},r_2}$.}
    \label{fig:electric_basis}
\end{figure}

The Kogut-Susskind Hamiltonian describing a lattice gauge theory with gauge group SU($N_c$) on a $d$ dimensional spatial lattice is given by
\begin{equation}
    \hat{H} = \frac{g^2}{2} \sum_l \hat{E}_l^2 - \frac{1}{2g^2} \sum_p \left(\hat{\Box}_p + \hat{\Box}_p^\dagger\right) \,,
\end{equation}
where $g$ is the gauge coupling, $\hat{E}_l^2$ is the electric energy operator on link $l$, and $\hat{\Box}_p$ is the trace of the product of parallel transporters around plaquette $p$. 
Once an orientation of the lattice has been chosen, the Hilbert space for this system is spanned by electric basis states. 
In them, an irreducible representation (irrep) $\mathcal{R}$ is attached to each link and a vector within said irrep is attached to each half-link, as shown in~\cref{fig:electric_basis}. In this work, we will orient horizontal links from left to right, and vertical links from bottom to top.
Plaquettes will be oriented counter-clockwise.
A more precise and pedagogical formulation of the Kogut-Susskind Hamiltonian and electric/magnetic bases can be found in the appendices of \cite{Burbano:2024uvn}.

Irreducible representations of SU($N_c$) can be specified by their Dynkin indices, 
\begin{align}
    \mathcal{R} \mathrel{\widehat{=}} (m_1,m_2,\cdots,m_{N_c-1})
    \,,
\end{align}
or by the corresponding Young diagram
\begin{align}
    (0,0,  \ldots, 0, 0) &\equiv \mathbf{1}\nonumber\\
    (1, 0, \ldots, 0, 0) &\equiv \wh\nonumber\\
    (0,0, \ldots, 0, 1) &\equiv \bl\nonumber\\
    (2,0, \ldots,0,0) &\equiv \whwhs\nonumber\\
    (0,0, \ldots, 0, 2) &\equiv \blbls\nonumber\\
    (0, 1,  \ldots,  0, 0) &\equiv \whwha\nonumber\\
    (0,\ldots, 0, 1, 0) &\equiv \blbla\nonumber\\
    (1,0, \ldots, 0,1) &\equiv \blwh\nonumber \,.
\end{align}
Note that the anti-fundamental representation will be represented by $\bl$ instead of a stack of $N_c-1$ boxes.
Components of the irrep $\mathcal{R}$, labeled by $r$, are represented by a Young tableaux that can be formed from the given Young diagram. 
The number of boxes in a given representation is given by 
\begin{align}
    n(\mathcal{R}) = \sum_{k=1}^{\lfloor N_c/2 \rfloor} k (m_k + m_{N-k}) + \frac{N_c}{2}m_{N_c/2}\,,
\end{align}
where the last term is only added for even $N_c$. 

\begin{figure}
    \centering
    \includegraphics[width=0.28\linewidth]{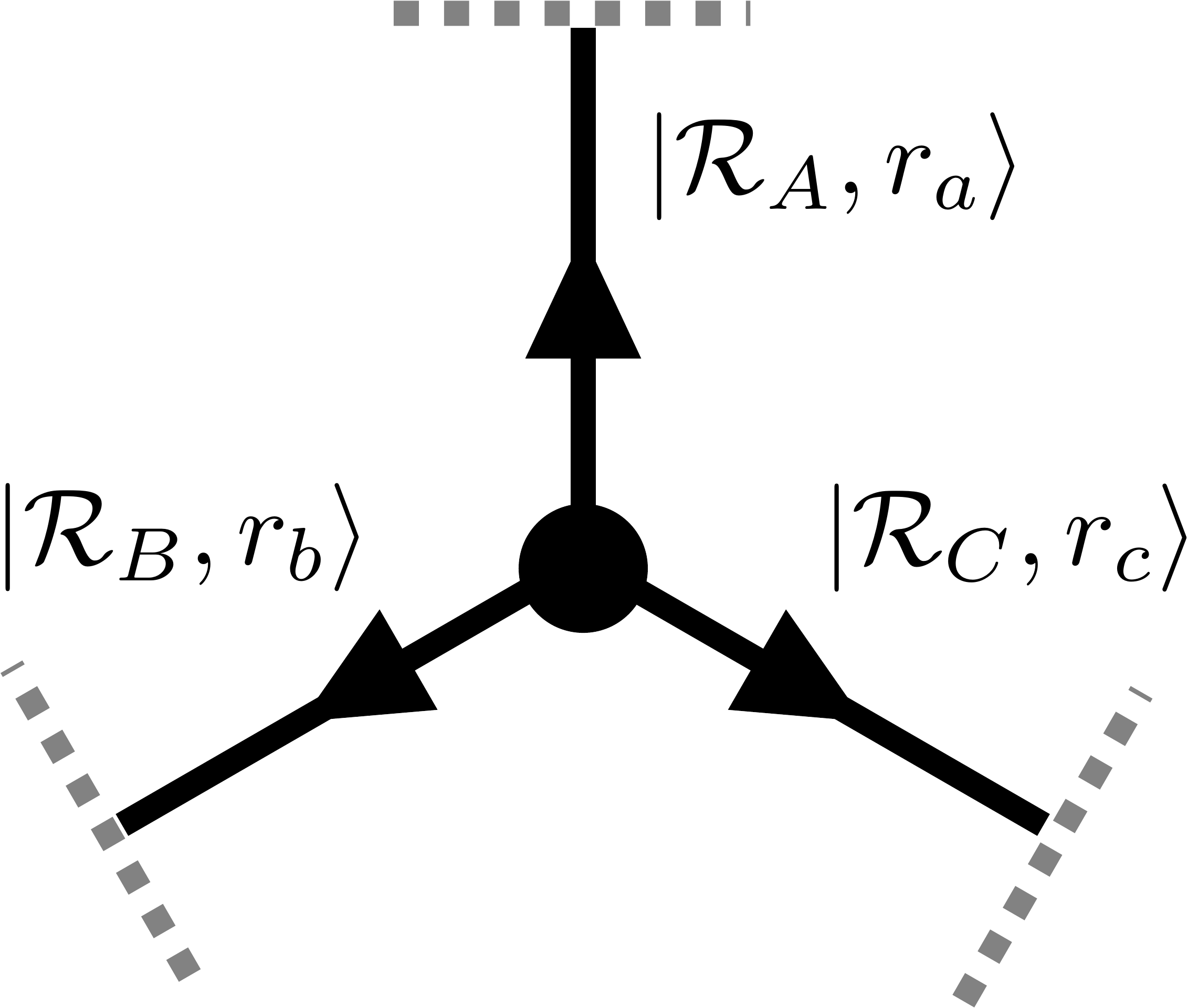}
    \caption{A generic vertex is shown in the electric basis. The linear combination \eqref{eq:vertex_factor} of these states yields the basis of~\cite{Ciavarella:2021nmj}.}
    \label{fig:electric_vertex}
\end{figure}

Gauss's law requires that at each vertex of the lattice the irrep components of all neighboring links combine together to form a singlet. 
It is well known that the contractions at vertices have to obey so-called Mandelstam constraints~\cite{Mandelstam:1978ed,Watson:1993zr}, which is usually implemented by using a trivalent lattice that only contains vertices with at most 3 links attached to them~\cite{Ciavarella:2021nmj,Zache:2023dko,Klco:2019evd,Raychowdhury:2019iki,Kadam:2022ipf}.
A general hypercubic lattice can be converted to such a trivalent lattice using the mechanism of point splitting~\cite{Burbano:2024uvn}.
The Gauss law constraints at each vertex can be enforced by requiring that the representations of all half-links emanating from a given vertex combine into singlets under the gauge group, forming vertex states
\begin{equation}
\label{eq:vertex_factor}
    \ket{v} \equiv \ket{\phi(\mathcal{R}_A,\mathcal{R}_B,\mathcal{R}_C)} = \sum_{r_a,r_b,r_c} \frac{C^{\Bar{\mathcal{R}_C},\Bar{r_c}}_{\mathcal{R}_A,r_a;\mathcal{R}_B,r_b}}{\sqrt{\text{dim}(\mathcal{R}_C)}} \ket{\mathcal{R}_A,r_a} \ket{\mathcal{R}_B,r_b} \ket{\mathcal{R}_C,r_c} \,,
\end{equation}
where $\mathcal{R}_A$, $\mathcal{R}_B$ and $\mathcal{R}_C$ are the representations on the links connecting to the trivalent vertex $v$, when we orient them outwards from the vertex as shown in~\cref{fig:electric_vertex} and $C^{\Bar{\mathcal{R}_C},\Bar{r_c}}_{\mathcal{R}_A,r_a;\mathcal{R}_B,r_b}$ is a SU($N_c$) Clebsch-Gordan coefficient. Note that in SU($N_c$), there may be multiple ways to form a representation, and the corresponding multiplicity index in the Clebsch-Gordan coefficient has been suppressed. The Hilbert space is then a subspace of the space spanned by the tensor product of all these vertex states
\begin{align}
    {\cal H} = {\rm span}\left( \bigotimes_v \ket{v} \right)
    \,.
\end{align}
The subspace corresponding to the physical Hilbert space has the gluing constraint (also referred to as an Abelian Gauss law) that half-links on a common link have the same representation.
This is the representation chosen in~\cite{Ciavarella:2021nmj}.

An alternative representation of the gauge invariant Hilbert space was introduced in~\cite{Ciavarella:2024fzw}, using the fact that SU($N_c$) representations can be visualized using an arrow notation. 
In this notation, $\wh$ and $\bl$ are represented by oriented lines with arrows indicating the orientation. By convention, we use an arrow compatible with the orientation of the lattice for the irrep $\wh$. Individual lines with arrows pointing in the same direction indicate that the corresponding boxes are arranged next to each other (in a symmetric fashion), while arrows going through multiple lines indicate that the corresponding boxes are arranged on top of each other (in an antisymmetric fashion), with examples given in~\cite{Ciavarella:2024fzw}. 
In this notation, the vertex factors in \cref{eq:vertex_factor} represent how the arrows of the three representations $\mathcal{R}_A,  \mathcal{R}_B,  \mathcal{R}_C$ combine or split at the given vertex, and gauge invariance requires conservation of oriented lines. 
A gauge invariant state in the absence of any charges therefore contains no source of oriented lines, and therefore all oriented lines form closed loops.
Such a state can therefore be represented by specifying all loops, and for links with multiple lines in the same direction how these lines combine into symmetric and antisymmetric pieces.
The Hilbert space is therefore given by
\begin{align}
    {\cal H} = {\rm span}\left( \ket{\{L_i, a_l\}}\right)
    \,,
\end{align}
where $L_i$ specifies the i'th loop, and $a_l$ how the arrows on link $l$ combine.
While this basis is highly non-local, it proves to be useful when performing an expansion in inverse powers of $1/N_c$, as will be discussed next.

\subsection{Simplifications from the \texorpdfstring{$1/N_c$}{1/Nc} expansion}

So far this discussion has been completely general, and did not rely on an expansion in $1/N_c$. 
Several important simplifications happen when performing this expansion, which we will now discuss.
First, to leading order in $1/N_c$ the Casimir of a given representation is given by
\begin{align}
\label{eq:casimir_leading}
    C(\mathcal{R}) &= n(\mathcal{R}) \frac{N_c}{2} \left( 1 + {\cal O}\left(1/N_c\right) \right) \,,
\end{align}
and the dimension of representations of SU($N_c$) scale with the number of boxes as
\begin{align}
    d(\mathcal{R}) \sim N_c^{n(\mathcal{R})} \,.
\end{align}

A second important simplification follows from the large $N_c$ scaling of matrix elements of the plaquette operators. Any physical gauge-invariant state can be built from the electric vacuum state by applying plaquette operators. This allows states to inherit a large $N_c$ suppression from the scaling of the plaquette matrix elements used to create the state. As shown in~\cite{Ciavarella:2024fzw}, at leading order in large $N_c$, the only unsuppressed states are those where loops enclose only a single plaquette. Working strictly to leading order in all terms would therefore require only specifying the number of boxes determining the representation at each plaquette, giving rise to a Hilbert space spanned by
\begin{align}\label{eq:trivial_hilbert}
    {\cal H} = {\rm span}\left(\ket{\{n(\mathcal{R}_p)\}}\right)\,.
\end{align}
The corresponding Hamiltonian is very simple, reflecting the well-known factorization of Wilson loops at large $N_c$ \cite{Makeenko:1980vm}, and as shown in~\cref{app:leading_Nc} is exactly solvable, leading to a theory with vanishing correlation length. 
Given that taking the continuum limit of the lattice theory requires tuning the lattice theory to the limit of infinite correlation length, one clearly needs to go beyond the strict $1/N_c$ limit.
In fact, in~\cite{Ciavarella:2024fzw} it was further shown that the presence of a loop suppresses the state by a factor proportional to $N_c^{1-m}$, with $m$ the number of plaquettes enclosed by the loop. 

There are several different effects at ${\cal O}(1/N_c)$ that need to be taken into account. 
First, both the electric and magnetic terms in the Hamiltonian receive corrections at that order. 
The Casimir of the different representations, which are required for the electric contribution to the Hamiltonian, need to be included at subleading order. 
The same is true for transitions between representations mediated by the magnetic contribution to the Hamiltonian, described by Clebsch-Gordan coefficients. 
Closed form, analytical expressions exist for the Casimirs of any SU($N_c$) group, and are summarized in~\cref{app:casimirs}. 
Such expressions do not exist for Clebsch-Gordan coefficients, and need to be worked out for the various transitions required. 
Some general relations that allow the relevant combinations of Clebsch-Gordan coefficients used in this work to be derived are given in~\cref{app:clebsch_gordan}.

Second, in order to obtain non-zero correlation lengths, working to ${\cal O}(1/N_c)$ requires introducing states where loops enclose $m=2$ plaquettes. 
This can be understood by the fact that each contraction of a fundamental and antifundamental index of a representation is suppressed by one power of $1/N_c$. 
Using the same argument, one can obtain a power counting in $1/N_c$ of matrix elements of the plaquette operator. 
Since a plaquette operator contains Wilson line operators in the fundamental representation on the four links enclosing the plaquette, a transition can create or annihilate up to 4 lines in the state it acts upon. 
Transitions in which fewer than four lines are created or annihilated are suppressed by powers of $1/N_c$, and the suppression is given by the difference from four of the number of lines annihilated or created. 
Examples of transitions allowed at this order are shown in~\cref{fig:transitions_1_1_1,fig:transitions_1_2_1,fig:transitions_1_2_2,fig:transitions_2_2_2}.

\section{Truncating the electric basis states}
In the discussion up to this point, we have retained the full Hilbert space of the gauge boson at each link.
Since this Hilbert space is infinite-dimensional, it has to be truncated to make numerical simulations feasible.
This truncation restricts the full Hilbert space to a subspace. In practice, one wishes to describe the expectation of observables in a low-energy subspace. For this reason, the electric basis truncation is formed from states exhibiting low energy density on the lattice.
Eigenstates of the electric Hamiltonian are characterized by the representations of the gauge fields on each link. 
Gauss' law, which has to be enforced on states in the physical Hilbert space, is most easily represented using such representations of the gauge field.
The most commonly used truncation of the Hilbert space is therefore based on the electric energy density, including only those states that have an electric energy density below a certain value in the lattice. 
This truncation can be shown to be very efficient in the strong coupling limit, while its efficiency at weaker coupling needs to be studied more carefully~\cite{Ciavarella:2021nmj,Ciavarella:2023mfc}. In the weak coupling limit, rigorous error bounds show there is an exponential convergence as the electric basis truncation is raised~\cite{Tong:2021rfv}.
In this work, we will choose to truncate based on the electric energy density and study the efficiency of this truncation by comparing results in the first few truncations.

The precise definition of the restriction of the energy density used can vary.
A common definition is to limit the electric energy on each link, which restricts the representations of the electric basis states at each link. 
Since the electric energy in a given link is given by the Casimir of the representation of the link, which as shown in~\cref{eq:casimir_leading} is determined only by the number of oriented lines on the link in the large $N_c$ limit, one can construct the subspace by including all gauge invariant states with at most $k$ oriented lines on each link. 
This corresponds to the $1/N_c$ limit of the basis used in~\cite{Ciavarella:2021nmj}. This basis includes all possible gauge-variant states. However, a large number of states are suppressed by large powers of $1/N_c$, as one can form arbitrarily large loops from the states contained in this subspace.
For this reason, it is useful to consider alternatives to this truncation that allow for an easier implementation of the constraints arising from the $1/N_c$ expansion.

An alternative to the previously considered construction~\cite{Ciavarella:2021nmj} of a truncated subspace in this work is inspired by the Krylov subspace construction. Krylov subspaces are spanned by the vectors generated by repeatedly applying an operator to an initial vector. These subspaces are the basis of the Lanczos algorithm and have proven to be useful in traditional lattice calculations~\cite{Wagman:2024rid,Hackett:2024xnx,Hackett:2024nbe}. In this work, a localized version of Krylov subspaces is used by constructing states through repeated applications of the plaquette operators, which make up the magnetic Hamiltonian on the electric vacuum state.
To limit the local energy densities that can be generated,  we restrict the number of times a plaquette operator can be applied in a given local region.
First, we require that we can apply a given plaquette operator at most $n_p$ times.
Since applications of neighboring plaquette operators both act on the shared link, we furthermore restrict the total number of link operators arising from these plaquette operators at each link to $n_l$.
The two numbers satisfy $n_l \geq n_p$, and the combination $(n_p, n_l)$ defines a plaquette-based truncation.

\begin{figure}
    \centering
    \includegraphics[width=0.5\linewidth]{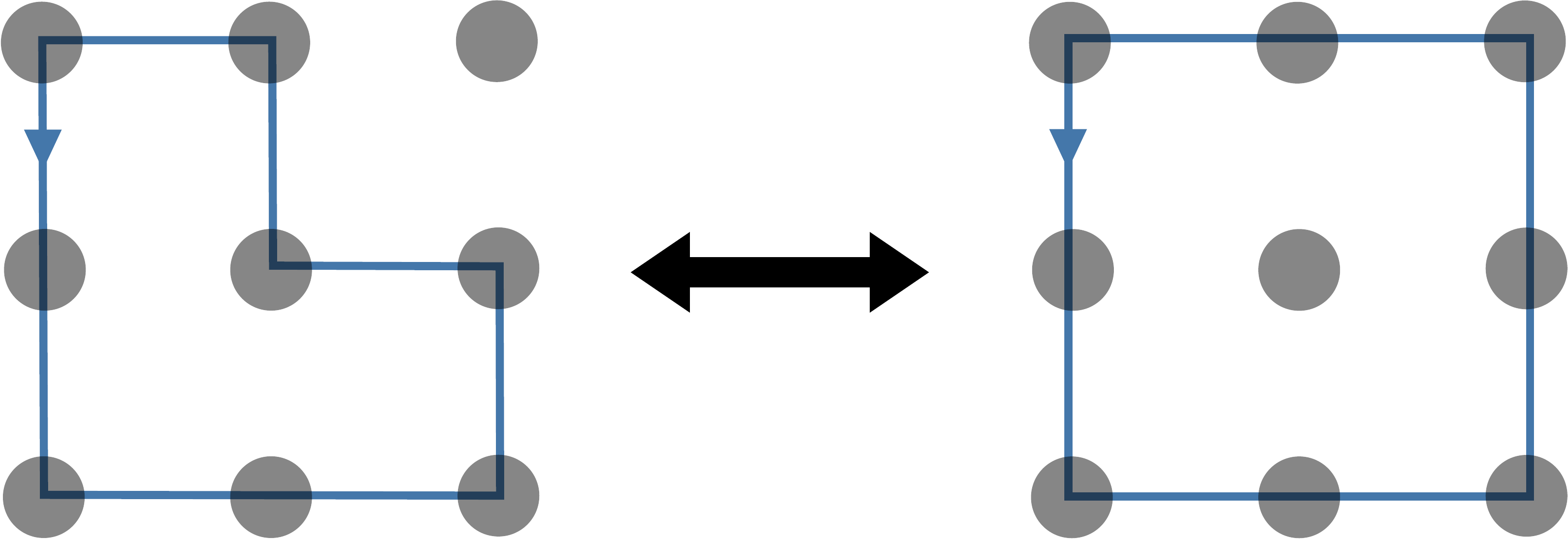}
    \caption{This transition is of $\order{1/N_c}$. This can be seen because it involves the deletion of a single joined line due to the action of a plaquette operator. However, creating this field configuration from the electric vacuum requires multiple $\order{1/N_c}$ transitions. Accordingly, in this work, we do not attempt to correctly include the matrix elements associated with this transition.}
    \label{fig:disallowed_transition}
\end{figure}

As is the case for any Krylov subspace construction, this latter construction is defined through the action of operators on sets of states that can be obtained through repeated applications on a starting state, in our case applying plaquette operators to the electric vacuum state. 
As discussed, the matrix elements of the plaquette operators depend on Clebsch-Gordan coefficients, which can be expanded in powers of $1/N_c$. 
The precise definition of the truncated subspace, therefore, depends on how this $1/N_c$ expansion is implemented.
In this work, we set all transitions between states that are suppressed by two or more powers of $1/N_c$ to zero, while keeping all transitions that arise at ${\cal O}(1)$ or  ${\cal O}(1/N_c)$. We also do not attempt to include transitions that only occur in states that are suppressed by multiple powers of $1/N_c$.
An example of a disallowed transition is given in~\cref{fig:disallowed_transition}.

Note that, due to transitions at $1/N_c$, the number of oriented lines on any given link can be smaller than the number of link operators applied. 
One can therefore add a restriction on the total number of oriented lines on each link of the states reached by the application of the plaquette operators, such that each truncation is labeled by the three numbers $(n_p, n_l, k)$.

The truncation introduced here implies a restriction on the local energy density on the lattice. 
By construction, all states in this truncation have irreps with at most $k$ oriented lines on each link. 
However, there are constraints on the local energy density in a 2-plaquette system as well.
Consider the total number of oriented lines in any adjacent pair of plaquettes,  arising from loops that are contained within the chosen pair. We specify this number, and label the largest possible number on the lattice by $\Lambda$.
The simplest $(1, 1, 1)$ truncation has at most a single plaquette in each pair excited, corresponding to $\Lambda = 4$. 
This is the truncation used in~\cite{Ciavarella:2024fzw}.
The $(1,2,1)$ truncation adds a single loop encircling both plaquettes, corresponding to $\Lambda = 6$.
The $(1,2,2)$ truncation adds states with one loop encircling each plaquette, therefore contributing at $\Lambda = 8$.
However, there are other states contributing to $\Lambda = 8$, which arise from the $(2,2,2)$ truncation. 
Thus, one might expect that the $(1,2,2)$ truncation might not be a systematic expansion, which will be confirmed in our numerical analysis later.

The next question is how to represent such truncations efficiently in a particular representation of the Hilbert space, and the states discussed so far allow for a particular convenient representation.
For the truncations up to $(2, 2, 2)$ that will be used in this work, states can be written as
\begin{equation}
    \ket{\{\mathcal{R}_p\},\{\mathcal{L}_l\}} = \mathcal{N}(\{\mathcal{R}_p\},\{\mathcal{L}_l\}) \prod_{l} \ket{\mathcal{L}_l}_l \bra{\mathcal{L}_l}_l \prod_{p} \hat{\Box}_p^{(\mathcal{R}_p)} \ket{0} \,,
    \label{eq:basisstate}
\end{equation}
where $\ket{0}$ is the electric vacuum, $\hat{\Box}_p^{(\mathcal{R}_p)}$ is the plaquette operator at plaquette $p$ with parallel transporters in the $\mathcal{R}_p$ representation, $\ket{\mathcal{L}_l}_l \bra{\mathcal{L}_l}_l$ is the projection of the state on link $l$ onto the representation $\mathcal{L}_l$ and $\mathcal{N}(\{\mathcal{R}_p\},\{\mathcal{L}_l\})$ is a normalization constant. 
Therefore, basis states are specified by assigning a representation to each plaquette and a representation to each link. The representations on the links are constrained to be an element of the Clebsch-Gordan decomposition of the representations on the neighboring plaquettes. Similar encodings of state labels onto plaquettes have been explored for U($1$)~\cite{Bauer:2021gek,Kane:2022ejm} and SU($2$) lattice gauge theories~\cite{ARahman:2022tkr,Mendicelli:2022ntz,Muller:2023nnk}. Note that the basis introduced in Ref.~\cite{Muller:2023nnk} was restricted to $2+1D$ honeycomb lattices with open boundary conditions due to the existence of redundant basis states.
In particular, unlike the cubic case, in these lattices all vertices are of the form shown in \cref{fig:electric_vertex}.
The truncation in $1/N_c$ in this work resolves this issue by removing the coupling to the redundant basis states. However, when the truncation is raised above $(2,2,2)$, the coupling to redundant basis states is reintroduced, causing this basis to break down.
Furthermore, all transitions required in these truncations can be represented by specifying the representations of all of the neighboring plaquettes, as well as the projection $\mathcal{L}_l$ on the links of the plaquette.

\subsection{($n_p=1$,$n_l=1$,$k=1$) (\texorpdfstring{$\Lambda = 4$}{Lambda = 4}) Truncation}

\begin{figure}
    \centering
    \includegraphics[width=0.7\linewidth]{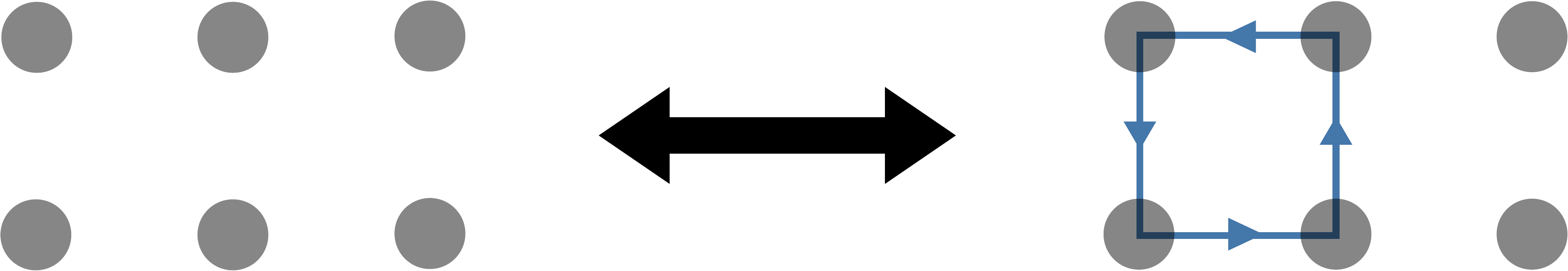}
    \caption{The transition allowed in the $(1,1,1)$ truncation.}
    \label{fig:transitions_1_1_1}
\end{figure}

The harshest truncation of the Hilbert space allows each plaquette operator to be applied only once, and each link to have only a single plaquette operator applied to it. At this truncation, once the representations on the plaquettes have been specified, the representations on the links will be as well. Therefore, this truncation requires a single qutrit per plaquette with basis states labeled by $\ket{\mathbf{1}}$, $\ket{\wh}$, and $\ket{\bl}$. This qutrit basis will contain unphysical states (such as neighboring plaquettes both in the $\ket{\wh}$ state) that do not correspond to a state in the gauge theory. However, the dynamics of the theory will not couple to these unphysical states. The existence of unphysical states will be a generic feature of the bases used in this work. With this basis, the truncated SU($N_c$) Hamiltonian at leading order in large $N_c$ is given by
\begin{align}
    \hat{H} & = \sum_p g^2 \left(N_c - \frac{1}{N_c}\right) \left(\ket{\wh}_p\bra{\wh}_p + \ket{\bl}_p\bra{\bl}_p\right)- \frac{1}{2g^2} \left(\prod_{\hat{n}} \ket{\mathbf{1}}_{p+\hat{n}} \bra{\mathbf{1}}_{p+\hat{n}} \right) \left(\ket{\wh}_p\bra{\mathbf{1}}_p + \ket{\bl}_p\bra{\mathbf{1}}_p + \text{h.c.}\right) \,,
\end{align}
where $p$ labels plaquettes on the lattice and $p+\hat{n}$ specifies neighboring plaquettes in the $\hat{n}$ direction. The transitions allowed here are represented in~\cref{fig:transitions_1_1_1}. Note that at this truncation, it is possible to enforce the $C$ symmetry (charge conjugation) locally and replace the qutrits with qubits. This is because at this truncation the operators $\hat{C}_p=(\ket{\wh}_p-\ket{\bl}_p)(\bra{\wh}_p-\bra{\bl}_p)$ commute with the Hamiltonian, indicating that states where a plaquette is in the state $\ket{\wh}_p-\ket{\bl}_p$ are decoupled from the electric vacuum at this truncation.
Explicitly, with the basis assignment
\begin{align}
   \ket{0}_p &= \ket{\mathbf{1}}_p \nonumber \\
   \ket{1}_p &= \frac{1}{\sqrt{2}} \left( \ket{\wh}_p + \ket{\bl}_p\right) \,,
\end{align}
the Hamiltonian for the even $C$ sector becomes
\begin{align}
    \hat{H} & = \sum_p g^2 \left(N_c - \frac{1}{N_c}\right) \ket{1}_p \bra{1}_p - \frac{1}{\sqrt{2}g^2} \left(\prod_{\hat{n}} \ket{0}_{p+\hat{n}} \bra{0}_{p+\hat{n}} \right) \hat{X}_p \,,
\end{align}
where $\hat{X}_p$ is the Pauli X operator acting on the qubit at plaquette $p$. Note that compared to our previous SU($3$) Hamiltonian~\cite{Ciavarella:2024fzw},  there is no term from the plaquette proportional to $\ket{1}_p \bra{1}_p$ as that was a unique feature of SU($3$). In the formulation in this work, this term will only be present in higher truncations of the theory.

\subsection{($n_p=1$,$n_l=2$,$k=1$) (\texorpdfstring{$\Lambda = 6$}{Lambda = 6}) Truncation}

\begin{figure}
    \centering
    \includegraphics[width=0.7\linewidth]{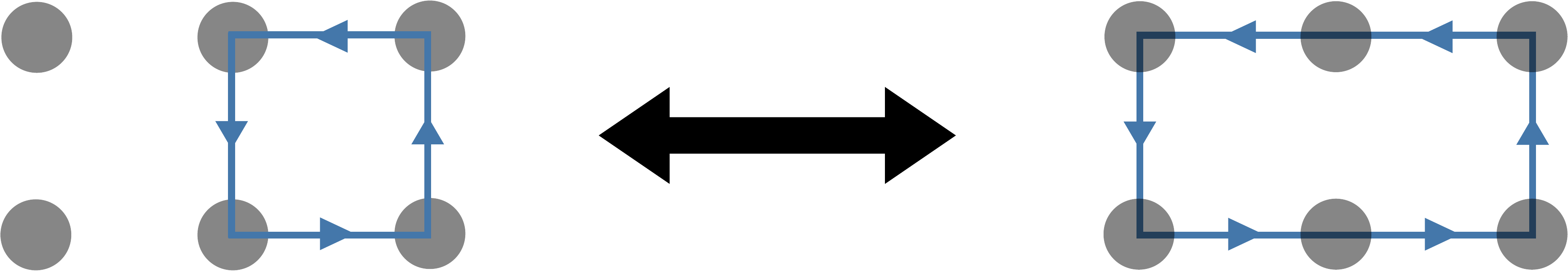}
    \caption{The new transition allowed in the $(1,2,1)$ truncation.}
    \label{fig:transitions_1_2_1}
\end{figure}

The next non-trivial truncation is the $(1,2,1)$ truncation. 
This truncation allows neighboring plaquettes to be excited with their shared link in the $\mathbf{1}$ representation. 
In a two plaquette system, there could be up to $\Lambda=6$ excited links. At leading order in large $N_c$, this is still equivalent to the $(1,1,1)$ truncation. 
At order $1/N_c$, the plaquette operator will be modified to allow neighboring plaquettes to be excited, as shown in \cref{fig:transitions_1_2_1}. In this truncation, states can still be represented using a qutrit per plaquette. 
If neighboring plaquettes are both in the $\ket{\bl}$ ($\ket{\wh}$) state it will be interpreted as a loop of electric flux flowing (counter)-clockwise around the two plaquettes with their shared link unexcited. 
The Hamiltonian is given by
\begin{align}
    \hat{H}^{(1,2,1),1/N_c} =& \hat{H}^{(1,2,1),1/N_c}_E + \hat{H}^{(1,2,1),1/N_c}_B \nonumber \\
    \hat{H}^{(1,2,1),1/N_c}_E = & g^2 \left(N_c - \frac{1}{N_c}\right) \sum_p  \Bigg[\left(\ket{\wh}_p\bra{\wh}_p + \ket{\bl}_p\bra{\bl}_p\right) \nonumber\\
    &- \frac{1}{4} \sum_{\hat{n}} \left( \ket{\wh}_p\bra{\wh}_p \ket{\wh}_{p+\hat{n}}\bra{\wh}_{p+\hat{n}} + \ket{\bl}_p\bra{\bl}_p \ket{\bl}_{p+\hat{n}}\bra{\bl}_{p+\hat{n}}  \right) \Bigg]\nonumber \\
     \hat{H}^{(1,2,1),1/N_c}_B =& -\frac{1}{2g^2} \sum_p \left(\prod_{\hat{n}} \ket{\mathbf{1}}_{p+\hat{n}} \bra{\mathbf{1}}_{p+\hat{n}} \right) \left(\ket{\wh}_p\bra{\mathbf{1}}_p + \ket{\bl}_p\bra{\mathbf{1}}_p + \text{h.c.}\right) \nonumber \\
     & -\frac{1}{2g^2N_c} \sum_p\sum_{\hat{k}} \ket{\wh}_{p+\hat{k}}\bra{\wh}_{p+\hat{k}} \prod_{\hat{n}\neq \hat{k}}\left(\ket{\mathbf{1}}_{p+\hat{n}} \bra{\mathbf{1}}_{p+\hat{n}}\right) \left(\ket{\wh}_p\bra{\mathbf{1}}_p + \text{h.c.}\right) \nonumber \\
     & -\frac{1}{2g^2N_c} \sum_p\sum_{\hat{k}} \ket{\bl}_{p+\hat{k}}\bra{\bl}_{p+\hat{k}} \prod_{\hat{n}\neq \hat{k}}\left(\ket{\mathbf{1}}_{p+\hat{n}} \bra{\mathbf{1}}_{p+\hat{n}}\right) \left(\ket{\bl}_p\bra{\mathbf{1}}_p + \text{h.c.}\right) \, .
     \label{eq:Ham121}
\end{align}
The form of the plaquette operator at this truncation can be derived using the methods in~\cref{app:clebsch_gordan}.
At this truncation, $\hat{C}_p$ is no longer a conserved quantity. However, it is still possible to construct a local basis for the even $C$ sector using a single qubit per plaquette. We can define (a non-local) loop basis by
\begin{equation}
    \ket{\{\mathcal{L}_i\}} = \mathcal{N}(\{\mathcal{L}_i\}) \left(\prod_i \prod_{p\in B_i} \ket{\wh}_p + \prod_i \prod_{p\in B_i} \ket{\bl}_p\right) 
\end{equation}
where $\{\mathcal{L}_i\}$ is a set of non-intersecting closed loops $\mathcal{L}_i$, $B_i$ is the minimal region of the lattice with $\partial B_i=\mathcal{L}_i$ and $\mathcal{N}(\{\mathcal{L}_i\})$ is a normalization constant. Any even $C$ state that can be generated through the dynamics of the electric vacuum or a state with localized excitations can be expressed in this basis. This loop basis can be represented locally by assigning a qubit to each plaquette. If the plaquette is an element of $B_i$ then the qubit is in the $\ket{1}$ state and otherwise the qubit is in the $\ket{0}$ state.
Regions of qubits in the $\ket{1}$ state correspond to there being an even superposition of flux flowing clockwise and counter-clockwise around the boundary of the region. With this encoding, the Hamiltonian is
\begin{align}
    \hat{H}^{(1,2,1)} =&  \sum_p g^2 (N_c - \frac{1}{N_c}) \ket{1}_p \bra{1}_p - \frac{g^2 (N_c - \frac{1}{N_c})}{4} \sum_{\hat{n}} \ket{1}_p \bra{1}_p \ket{1}_{p+\hat{n}} \bra{1}_{p+\hat{n}} \nonumber \\
    & - \frac{1}{\sqrt{2}g^2} \left(\prod_{\hat{n}} \ket{0}_{p+\hat{n}} \bra{0}_{p+\hat{n}} \right) \hat{X}_p  -\frac{1}{2g^2N_c} \sum_{\hat{k}} \ket{1}_{p+\hat{k}} \bra{1}_{p+\hat{k}} \left(\prod_{\hat{n}\neq \hat{k}} \ket{0}_{p+\hat{n}} \bra{0}_{p+\hat{n}} \right) \hat{X}_p \,.
\end{align}

\subsection{($n_p=1$,$n_l=2$,$k=2$) Truncation}

\begin{figure}
    \centering
    \includegraphics[width=0.7\linewidth]{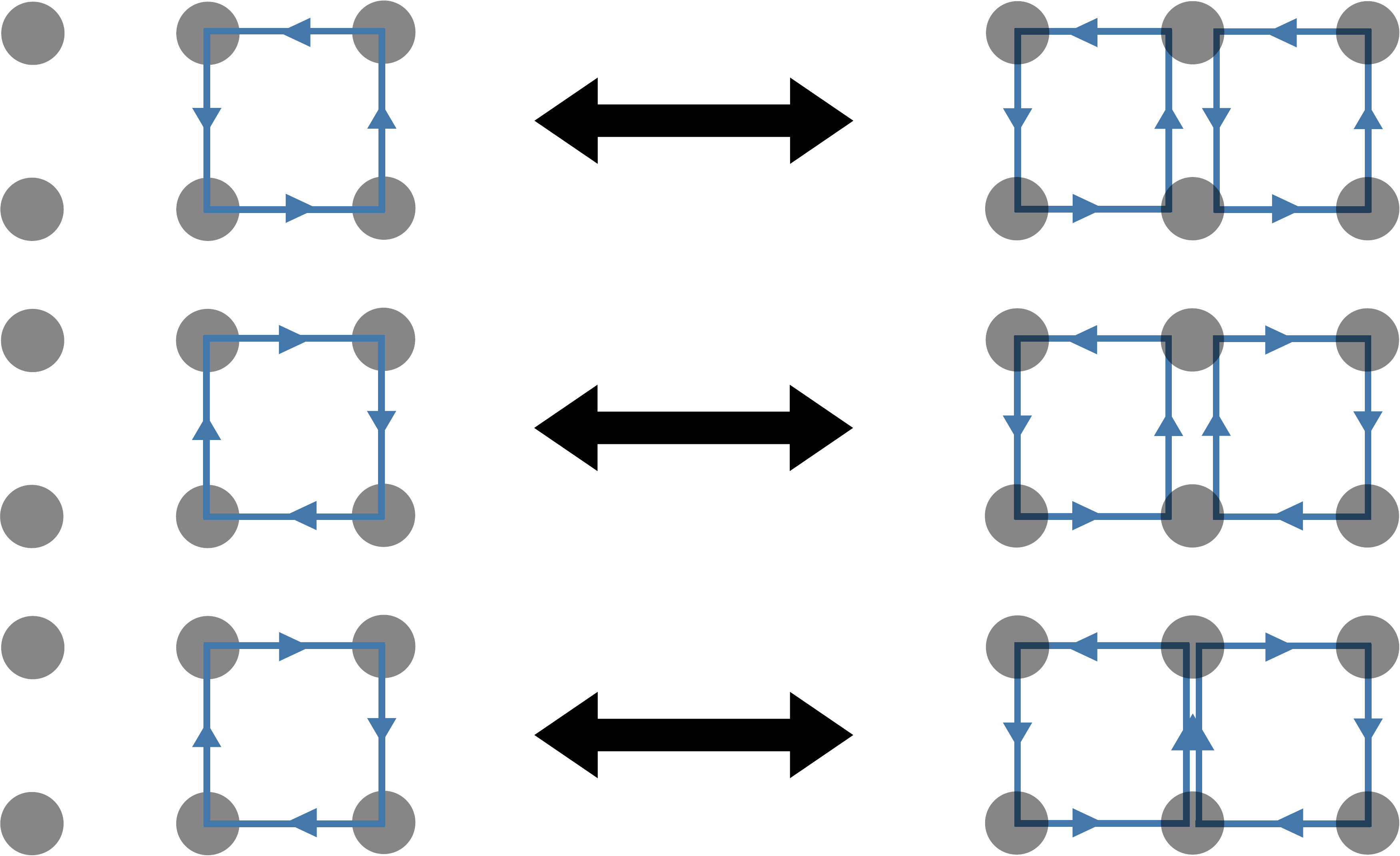}
    \caption{New transitions allowed in the $(1,2,2)$ truncation. The state depicted on the top right corresponds to a state of the form $\ket{\wh}_p\ket{1}_l\ket{\wh}_{p+\hat{n}}$. From \eqref{eq:plaq_elements}, we see this transition has a prefactor $\sqrt{1-\frac{1}{N_c^2}}$, once we recall that the adjoint representation $\blwh$ has dimension $N_c^2-1$. Similarly, the state on the middle right corresponds to $\ket{\wh}_p\ket{1}_l\ket{\bl}_{p+\hat{n}}$, and the associated transition has prefactor $\sqrt{\frac{1}{2}\qty(1+\frac{1}{N_c})}$, once we recall that $\whwhs$ has dimension $N_c(N_c + 1)/2$. Finally, the state on the lower right corresponds to $\ket{\wh}_p\ket{0}_l\ket{\bl}_{p+\hat{n}}$, and the associated transition has prefactor $\sqrt{\frac{1}{2}\qty(1-\frac{1}{N_c})}$, once we recall that $\whwhs$ has dimension $N_c(N_c - 1)/2$.}
    \label{fig:transitions_1_2_2}
\end{figure}

\label{sec:truncate12}
The next $(n_p,n_l,k)$ truncation will still allow each plaquette operator to only be applied a single time, but will allow neighboring plaquette operators to be applied and excite higher irreps. 
At leading order in large $N_c$, these shared links can support higher irreps labeled by $\blwh$ (the adjoint irrep), $\whwhs$ (two indices combined symmetrically), and $\whwha$ (two indices combined anti-symmetrically), along with their conjugate representations. 
Note that for SU($3$) $\bl$ is the same as the $\whwha$ representation, but for larger $N_c$ this is a different representation. 
To represent electric basis states, a qutrit will be assigned to each plaquette, and a qubit will be assigned to each link. 
When all qubits are in the zero state, the qutrit basis assignment is the same as the previous section, specifying if there is a loop of electric flux flowing around the plaquette. 
When neighboring plaquettes are excited, there is an ambiguity in the state of their shared link. 
If both plaquettes have a loop of flux in the same orientation, their shared link can be in the $\mathbf{1}$ or $\blwh$ irrep. 
However, at leading order in large $N_c$ the $\mathbf{1}$ irrep is not present. 
If the plaquettes have opposite orientations, the shared link can be in the $\whwhs$ or $\whwha$, or the conjugate of these representations, depending on the orientation of the link. The qubit on the shared link will be used to resolve this ambiguity.
If the qubit is in the $\ket{0}$ state, the shared link will be in the $\mathbf{1}$ or $\whwha$ depending on the neighboring qutrits and if the qubit is in the $\ket{1}$ state the shared link will be in the $\mathbf{\blwh}$ or $\mathbf{\whwhs}$ irrep. 
With this basis, the SU($N_c$) Hamiltonian at this truncation on a two-dimensional lattice is given by
\begin{align}
    \hat{H}^{(1,2,2)} = & \hat{H}^{(1,2,2)}_E + \hat{H}^{(1,2,2)}_B \nonumber \\
    \hat{H}^{(1,2,2)}_E = & g^2 \sum_p \left(N_c - \frac{1}{N_c}\right) \left(\ket{\wh}_p\bra{\wh}_p + \ket{\bl}_p\bra{\bl}_p\right) \nonumber \\
    &+ g^2 \sum_{\langle p,p'\rangle} \Bigg[ -\frac{N_c - \frac{1}{N_c}}{2} \left(\ket{\wh}_{p}\bra{\wh}_p \ket{\wh}_{p'}\bra{\wh}_{p'} + \ket{\bl}_{p}\bra{\bl}_p \ket{\bl}_{p'}\bra{\bl}_{p'}\right) \ket{0}_l\bra{0}_l \nonumber \\
    & + \frac{1}{2N_c} \left(\ket{\wh}_{p}\bra{\wh}_p \ket{\wh}_{p'}\bra{\wh}_{p'}
    + \ket{\bl}_{p}\bra{\bl}_p \ket{\bl}_{p'}\bra{\bl}_{p'}\right) \ket{1}_l\bra{1}_l \nonumber \\
    & - \frac{1 + \frac{1}{N_c}}{2}\left(\ket{\wh}_{p}\bra{\wh}_p \ket{\bl}_{p'}\bra{\bl}_{p'}
    + \ket{\bl}_{p}\bra{\bl}_p \ket{\wh}_{p'}\bra{\wh}_{p'}\right) \ket{0}_l\bra{0}_l  \nonumber \\
    & + \frac{1 - \frac{1}{N_c}}{2}\left(\ket{\wh}_{p}\bra{\wh}_p \ket{\bl}_{p'}\bra{\bl}_{p'}
    + \ket{\bl}_{p}\bra{\bl}_p \ket{\wh}_{p'}\bra{\wh}_{p'}\right) \ket{1}_l\bra{1}_l \Bigg] \nonumber \\
    \hat{H}^{(1,2,2)}_B = & -\frac{1}{2g^2} \sum_p \left( \prod_{\hat{n}} \hat{C}^{p+\hat{n}}_\wh \ket{\wh}_p \bra{\mathbf{1}}_p + \prod_{\hat{n}} \hat{C}^{p+\hat{n}}_\bl \ket{\bl}_p \bra{\mathbf{1}}_p + \text{h.c.}\right) \nonumber \\
    \hat{C}^{p+\hat{n}}_\mathbf{R} = &\ket{\mathbf{1}}_{p+\hat{n}} \bra{\mathbf{1}}_{p+\hat{n}} \ket{0}_l \bra{0}_l + \sqrt{1 - \frac{1}{N_c^2}} \ket{\mathbf{R}}_{p+\hat{n}} \bra{\mathbf{R}}_{p+\hat{n}} \ket{1}_l \bra{0}_l \nonumber \\
    & + \sqrt{\frac{1}{2}\left(1 - \frac{1}{N_c}\right)} \ket{\mathbf{\Bar{R}}}_{p+\hat{n}} \bra{\mathbf{\Bar{R}}}_{p+\hat{n}} \ket{0}_l \bra{0}_l + \sqrt{\frac{1}{2}\left(1 + \frac{1}{N_c}\right)} \ket{\mathbf{\Bar{R}}}_{p+\hat{n}} \bra{\mathbf{\Bar{R}}}_{p+\hat{n}} \ket{1}_l \bra{0}_l \,,
    \label{eq:Ham12}
\end{align}
where $p$ and $p'$ are neighboring plaquettes and $l$ is their shared link. Note that the matrix elements of the plaquette operator at this truncation can be derived using the results from the authors' previous work~\cite{Ciavarella:2024fzw} as described in~\cref{app:clebsch_gordan}. The new transitions included in this truncation are depicted in~\cref{fig:transitions_1_2_2}. The Hamiltonian in higher spatial dimensions takes a similar form, except the plaquette operators have additional projectors to ensure no link carries more than two excitations.

The $1/N_c$ corrections at this truncation require including states with neighboring plaquettes excited and their shared link in the $\mathbf{1}$ irrep. In the loop notation, these states look like loops of electric flux extending over multiple plaquettes. At this truncation, these states can still be represented using the basis described in~\cref{sec:truncate12}. The Hamiltonian becomes
\begin{align}
    \hat{H} &= \hat{H}^{(1,2,2)}_E + \hat{H}^{(1,2,2),1/N_c}_B \nonumber \\
    \hat{H}^{(1,2,2),1/N_c}_B &= \hat{H}^{(1,2,2)}_B - \frac{1}{2g^2 N_c} \sum_p\sum_{\hat{k}} \left( \ket{\wh}_{p+\hat{k}} \bra{\wh}_{p+\hat{k}} \ket{0}_l \bra{0}_l \prod_{\hat{n}\neq \hat{k}}\hat{C}^{p+\hat{n}}_{\wh} \ket{\wh}_p \bra{\mathbf{1}}_p +\text{h.c.} \right) \nonumber \\
    & - \frac{1}{2g^2 N_c} \sum_p\sum_{\hat{k}} \left( \ket{\bl}_{p+\hat{k}} \bra{\bl}_{p+\hat{k}} \ket{0}_l \bra{0}_l \prod_{\hat{n}\neq \hat{k}}\hat{C}^{p+\hat{n}}_{\bl} \ket{\bl}_p \bra{\mathbf{1}}_p +\text{h.c.} \right) \,,
\end{align}
where $\hat{k}$ is a unit vector in the $k$-th direction, and $l$ is the shared link for plaquettes $p$ and $p+\hat{k}$. At this truncation, it is still possible to construct a loop basis for the even $C$ sector similar to the one used for the $(1,2,1)$ truncation. However, at this truncation, it is possible for disjoint loops to share links, which prevents an analogous construction of a local basis for the even $C$ sector.

\subsection{($n_p=2$,$n_l=2$,$k=2$) (\texorpdfstring{$\Lambda = 8$}{Lambda = 8}) Truncation}
\begin{figure}
    \centering
    \includegraphics[width=0.7\linewidth]{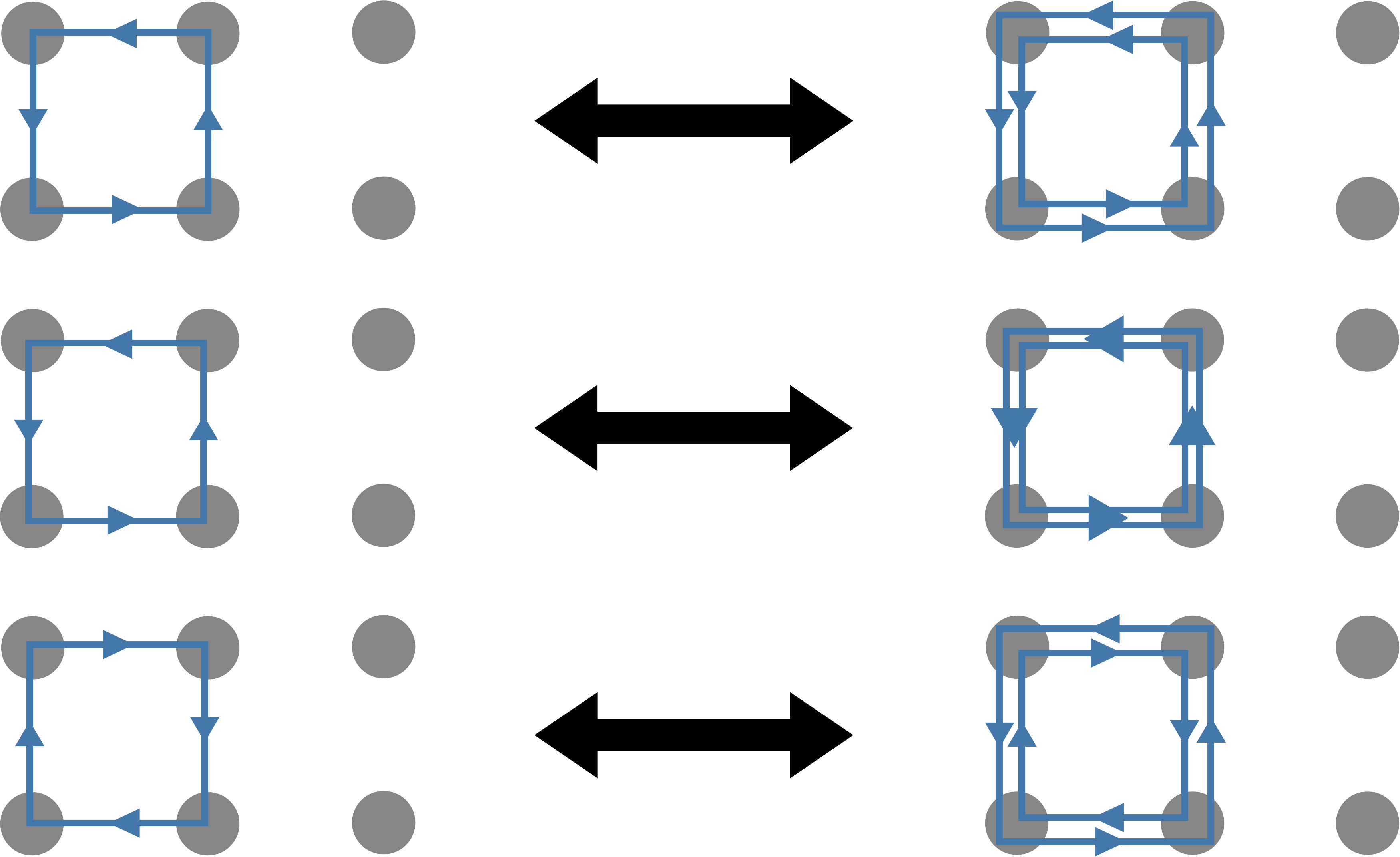}
    \caption{New transitions allowed in the $(2,2,2)$ truncation.}
    \label{fig:transitions_2_2_2}
\end{figure}

The next truncation allows each plaquette operator to be applied twice and each link to have at most two plaquette operators applied to it. In a two plaquette system, this would allow up to $\Lambda=8$ units of electric flux. This will add states that have plaquettes with a loop of flux in the $\whwha$, $\blbla$, $\blwh$, $\whwhs$, or $\blbla$ representation, as shown in \cref{fig:transitions_2_2_2}. Basis states at this truncation can be represented similarly to the (1,2,2) truncation, with the qutrits replaced by a qu$8$. The qubits on the links will be used to resolve ambiguities as before.
The Hamiltonian at leading order in large $N_c$ is given by
\begin{align}
    \hat{H}^{(2,2,2)} & = \hat{H}^{(2,2,2)}_E + \hat{H}^{(2,2,2)}_B \nonumber \\
    \hat{H}^{(2,2,2)}_E & = \hat{H}^{(1,2,2)}_E + \sum_p 2 N_c g^2 \ket{\blwh}_p \bra{\blwh}_p + 2g^2 \left(N_c+1 - \frac{2}{N_c} \right) \left( \ket{\whwhs}_p \bra{\whwhs}_p +\ket{\blbls}_p \bra{\blbls}_p \right) \nonumber \\
    & + 2g^2 \left(N_c-1 - \frac{2}{N_c}\right) \left(\ket{\whwha}_p \bra{\whwha}_p +\ket{\blbla}_p \bra{\blbla}_p \right) \nonumber \\
    \hat{H}^{(2,2,2)}_B & = \hat{H}^{(1,2,2)}_B -\frac{1}{2g^2} \sum_p \prod_{\hat{n}} \ket{\mathbf{1}}_{p+\hat{n}} \bra{\mathbf{1}}_{p+\hat{n}} \biggl( \ket{\whwha}_p \bra{\wh}_p + \ket{\blbla}_p \bra{\bl}_p + \ket{\blwh}_p \bra{\wh}_p + \ket{\blwh}_p \bra{\bl}_p \nonumber \\
    & + \ket{\whwhs}_p \bra{\wh}_p + \ket{\blbls}_p \bra{\bl}_p + \text{h.c.} \biggr) \,.
\end{align}
The $1/N_c$ corrections to this Hamiltonian are obtained by replacing $\hat{H}^{(1,2,2)}_B$ with $\hat{H}^{(1,2,2), 1/N_c}_B$.

In the case of SU($3$), $\whwha=\bl$ and $\blbla=\wh$, which means one only needs a qu$6$ for each plaquette instead of a qu$8$. Explicitly for SU($3$), the Hamiltonian at this truncation is 
\begin{align}
    \hat{H}^{(2,2,2)} & = \hat{H}^{(2,2,2)}_E + \hat{H}^{(2,2,2)}_B \nonumber \\
    \hat{H}^{(2,2,2)}_E & = \hat{H}^{(1,2,2)}_E + \sum_p 6 g^2 \ket{\blwh}_p \bra{\blwh}_p + g^2 \frac{20}{3}\left( \ket{\whwhs}_p \bra{\whwhs}_p +\ket{\blbls}_p \bra{\blbls}_p \right) \nonumber \\
    \hat{H}^{(2,2,2)}_B & = \hat{H}^{(1,2,2)}_B -\frac{1}{2g^2} \sum_p \prod_{\hat{n}} \ket{\mathbf{1}}_{p+\hat{n}} \bra{\mathbf{1}}_{p+\hat{n}} \biggl( \ket{\bl}_p \bra{\wh}_p + \ket{\blwh}_p \bra{\wh}_p + \ket{\blwh}_p \bra{\bl}_p \nonumber \\
    & + \ket{\whwhs}_p \bra{\wh}_p + \ket{\blbls}_p \bra{\bl}_p + \text{h.c.} \biggr) \,.
\end{align}

\section{Numerical Comparison}
\begin{figure}
    \centering
    \includegraphics[width=0.55\linewidth]{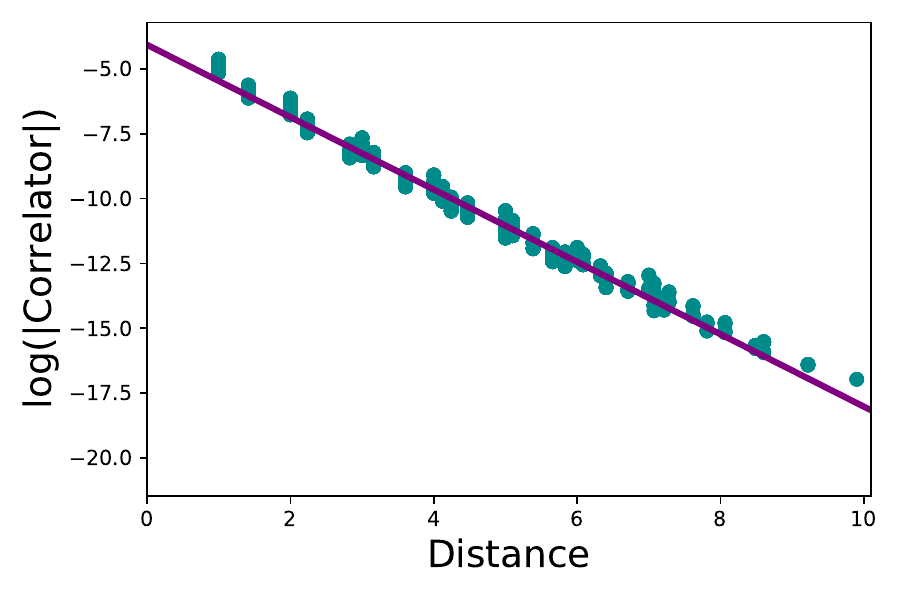}
    \caption{The blue points show the correlation function defined in Eq.~\eqref{eq:CorrFn} for the $(1,1,1)$ truncation with $g=0.8$ on an $8\times8$ lattice. The purple line is the fit to an exponential function.}
    \label{fig:CorrFn}
\end{figure}

\begin{figure}
    \centering
    \includegraphics[width=0.8\linewidth]{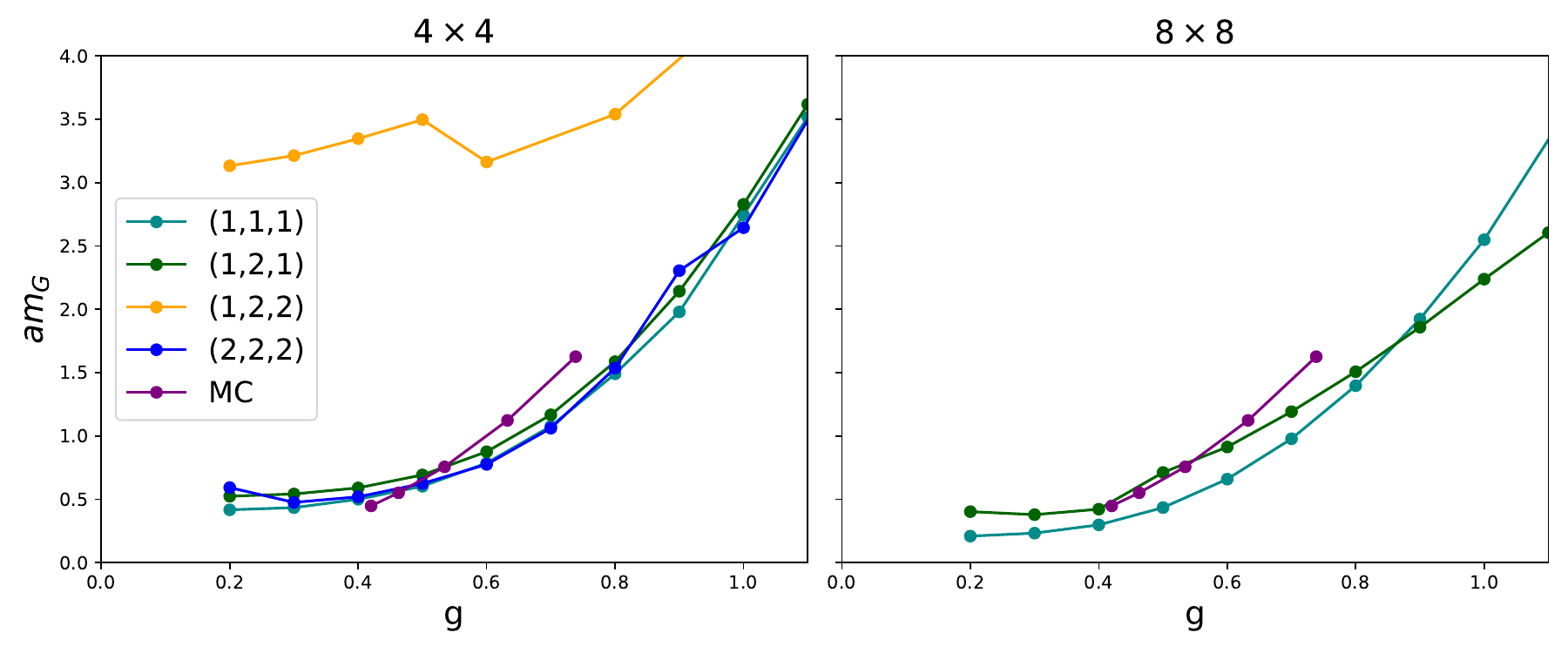}
    \caption{Mass of the $0^{++}$ glueball computed at different truncations as a function of $g$. The left panel shows the DMRG calculations performed on a $4\times4$ lattice and the right panel shows the DMRG calculations performed on an $8\times8$ lattice. The Monte Carlo data was taken from Table XVIII of reference~\cite{Teper:1998te}.}
    \label{fig:gluemass}
\end{figure}

\FloatBarrier
To understand how these truncations perform in practice, the ground states of these SU($3$) Hamiltonians were computed for small lattices in $2+1D$ using the density matrix renormalization group (DMRG)~\cite{White:1992zz} implemented using the C++ and Julia {\tt iTensor} libraries~\cite{Fishman_2022}. These calculations were performed using open boundary conditions. 
Calculations were performed for each truncation including the $1/N_c$ corrections. Details of the DMRG are given in~\cref{app:DMRG}.
Since the continuum limit is located at $g=0$, higher truncations should allow a closer approach to the continuum limit, and as the truncation is raised, the truncated Hamiltonian should approximate the untruncated theory better at small couplings $g$. 
To understand how small a lattice spacing can be achieved, the mass of the $0^{++}$ glueball was computed. 
In $3+1D$, the mass could be matched to a traditional lattice calculation of the glueball mass to set the lattice spacing. 
In traditional lattice calculations, particle masses are extracted by fitting correlation functions of operators with the appropriate quantum numbers. 
In this work, the glueball mass in $2+1D$ will be computed by fitting the vacuum correlation function of $\hat{\mathcal{O}}(p) = \ket{\wh}_p\bra{\wh}_p + \ket{\bl}_p\bra{\bl}_p$ to an exponential. Explicitly, the set of data points used in the fit was
\begin{equation}
    \left\{\left( \norm{\Vec{r}_1 - \Vec{r}_2}, \bra{\Omega}\hat{\mathcal{O}}(\Vec{r}_1) \hat{\mathcal{O}}(\Vec{r}_2) \ket{\Omega} - \bra{\Omega}\hat{\mathcal{O}}(\Vec{r}_1) \ket{\Omega} \bra{\Omega}\hat{\mathcal{O}}(\Vec{r}_2) \ket{\Omega} \right): \norm{\Vec{r}_1 - \Vec{r}_2} \geq 1 \right\} \,,
    \label{eq:CorrFn}
\end{equation}
where $\ket{\Omega}$ is the vacuum state found using DMRG. For points that are a distance $d \equiv \norm{\Vec{r}_1 - \Vec{r}_2}$ apart, this correlation function asymptotes to $a e^{-m_G d}$, which allows one to extract the glueball mass, $m_G$. An example of a correlation function is shown in~\cref{fig:CorrFn} for the $(1,1,1)$ truncation with $g=0.8$ on an $8\times8$ lattice. The points at fixed separations are not the same due to the use of the open boundary conditions in a finite volume. It is interesting to note that the points at non-integer separations (corresponding to lattice sites connected by a diagonal line) are still along the exponential fit. This is a reflection of the restoration of rotational symmetry, which is necessary to approach the continuum limit. The results of this calculation for all truncations are shown in~\cref{fig:gluemass} alongside a calculation of the glueball mass performed with the traditional action~\cite{Teper:1998te}.
Surprisingly, these low-lying truncations (with the exception of the $(1,2,2)$ truncation) appear to be consistent with the traditional action to around $g \approx 0.4$. 
These calculations were done on different lattice sizes, so exact agreement would not be expected due to finite volume effects. Finite volume effects are not expected to be large in this calculation as the finite volume corrections to a mass scale as $\mathcal{O}(e^{-mL})$~\cite{Luscher:1985dn}.

The $(1,2,2)$ truncation fails to agree with the untruncated calculation at all and has the lattice spacing freeze out at a larger value than the other truncations. 
At first sight, this is unexpected as the $(1,2,2)$ truncation has more states included than the $(1,2,1)$ truncation, so it would be expected to be in better agreement with the untruncated theory. 
It is unclear if similar behavior will occur in truncations with larger $(n_P,n_L,k)$ where $n_P \neq n_L$. 
Note that a strong coupling expansion of the vacuum state performed with the $(1,1,1)$ truncation will agree with an expansion performed with the untruncated Hamiltonian to first order. 
The $(2,2,2)$ truncation agrees to second order while the $(1,2,2)$ truncation is only in agreement to first order. 
Naively, improving the performance of the truncation of the theory at weak coupling would also improve agreement in the strong coupling regime. 
From this perspective, it would not be unexpected for the $(1,2,2)$ truncation to fail to improve over the $(1,1,1)$ truncation. However, this does not explain why it appears to perform worse than the $(1,1,1)$ truncation.

\begin{figure}
    \centering
    \includegraphics[width=0.8\linewidth]{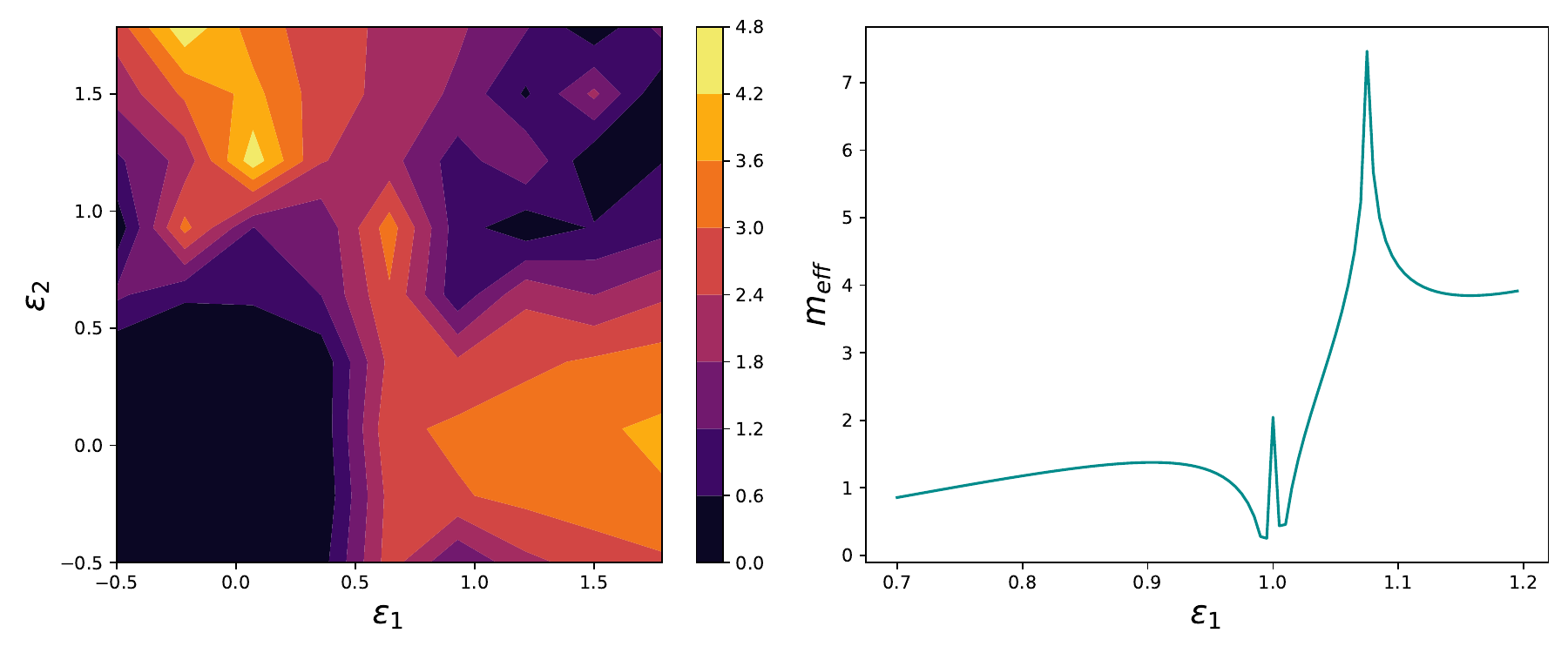}
    \caption{Effective mass computed on a $5\times1$ lattice with $g=0.5$. The left panel shows the effective mass as a function of $\epsilon_1$ and $\epsilon_2$. The right panel shows the effective mass as a function of $\epsilon_1$ with $\epsilon_2=0$.}
    \label{fig:MassChain}
\end{figure}

\begin{figure}
    \centering
    \includegraphics[width=0.55\linewidth]{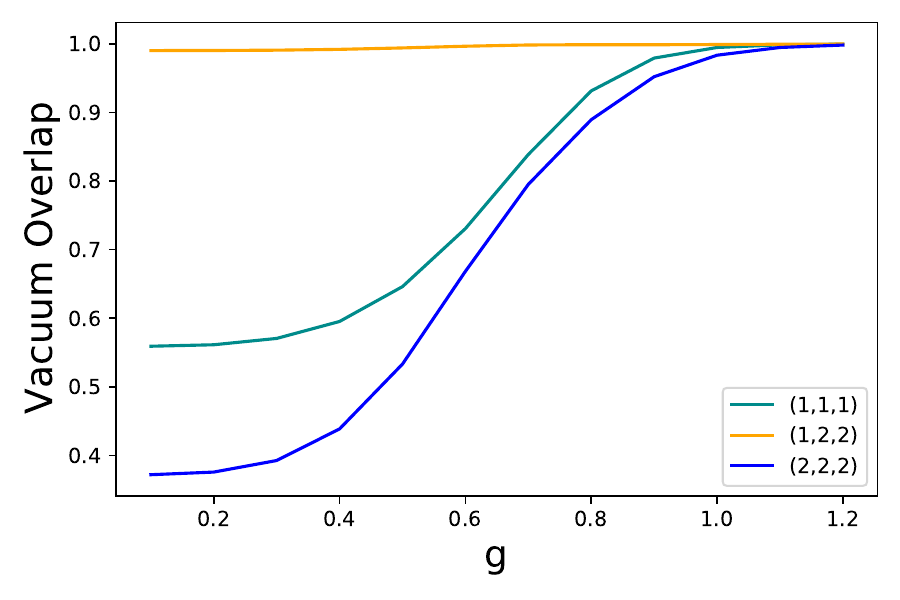}
    \caption{Overlap of the vacuum state with the vacuum state of the zero correlation Hamiltonian as a function of $g$ for a plaquette chain of length $5$ with PBC.}
    \label{fig:largeNoverlap}
\end{figure}

The unexpected behavior of the $(1,2,2)$ truncation can be better understood by taking a more general view of Hamiltonian truncations. Instead of viewing truncations as a projection on the Hilbert space, they can be treated as a deformation of the untruncated Hamiltonian that decouples parts of the Hilbert space. Explicitly, a deformed version of the Kogut-Susskind Hamiltonian can be given by
\begin{equation}
    \hat{H}(\Vec{\epsilon}) = \frac{g^2}{2} \hat{H}_E + \frac{1}{2g^2}\sum_i \epsilon_i \hat{H}_i\,,
\end{equation}
where $\hat{H}_E$ is the full electric Hamiltonian and $\hat{H}_i$ are different partitions of the magnetic Hamiltonian. The $\hat{H}_i$ are chosen so that the $n$-th truncation corresponds to $\epsilon_i=1$ for $i\leq n$ and $\epsilon_i = 0$ for $i>n$. Generically, $\Vec{\epsilon}$ specifies some location in theory space with specific points corresponding to various truncations and the untruncated Hamiltonian. The untruncated Hamiltonian is given by all $\epsilon_i=1$, and as $g\rightarrow0$ there is a critical point corresponding to continuum Yang Mills. Ultimately, one would like to perform a lattice simulation in the neighborhood of this critical point. However, it may be the case that certain points in theory space have behavior that takes them farther away from this critical point than one may initially expect. To show that this is the case for the $(1,2,2)$ truncation, a numerical simulation of 
\begin{equation}
    \hat{H}(\Vec{\epsilon}) = \hat{H}^{(1,1,1)} + \epsilon_1 \left(\hat{H}^{(1,2,2)} - \hat{H}^{(1,1,1)}\right) + \epsilon_2 \left(\hat{H}^{(2,2,2)} - \hat{H}^{(1,2,2)}\right)
\end{equation}
was performed for a $5\times1$ lattice with PBC. The vacuum state was computed using exact diagonalization. This system size was too small to extract a correlation length by fitting, so an effective mass was computed instead. Explicitly, the correlation function $G(x) = \bra{\psi}\hat{\mathcal{O}}(x) \hat{\mathcal{O}}(0) \ket{\psi} -\bra{\psi}\hat{\mathcal{O}}(0) \ket{\psi}^2$ was computed where $\ket{\psi}$ is the vacuum state. 
The correlations in the system were estimated from the effective mass $m_{eff}=-\log\left(\abs{\frac{G(2)}{G(1)}}\right)$. The effective mass as a function of $\epsilon_1$ and $\epsilon_2$ for $g=0.5$ is shown in~\cref{fig:MassChain}.
These calculations show that the $(1,2,2)$ truncation is close to a point in theory space where the effective mass is divergent. \Cref{app:12Decouple} shows how dropping all $1/N_c$ terms from the electric energy and plaquette operator in the $(1,2,2)$ truncation gives a Hamiltonian with zero correlation length for any coupling $g$. 
As $g\rightarrow0$, the effective mass should go to zero in the untruncated theory. The failure of the effective mass to decrease indicates that this is not a useful truncation of the Kogut-Susskind Hamiltonian. To verify that this is only a feature of the $(1,2,2)$ truncation, the vacuum was computed for a plaquette chain of length $5$ with PBC. The overlap of the vacuum state for each truncation with the vacuum state of the zero correlation length Hamiltonian in~\Cref{app:12Decouple} as a function of $g$ is shown in~\cref{fig:largeNoverlap}. As this figure shows, only the $(1,2,2)$ has a large overlap with the zero correlation truncation for all values of $g$. This indicates that only the $(1,2,2)$ truncation is perturbatively close to this limit for all values of $g$.

We finally consider how these truncated Hamiltonians can be used to perform an actual calculation. 
One may expect that simulations with different truncations would be performed to estimate errors due to truncating the Hamiltonian. 
However, this is not necessary, as ultimately we are interested in deviations between our truncated Hamiltonian and the continuum limit, not deviations from the untruncated Kogut-Susskind Hamiltonian. 
The truncations of the Hamiltonian performed here are compatible with all of the symmetries of the Hamiltonian and preserve locality. 
This means that our truncated Hamiltonian can be described by a Symanzik effective action whose leading order piece is the Yang-Mills action plus irrelevant operators multiplied by powers of the lattice spacing. 
There are no additional relevant operators in the action, as any other relevant operator would break the symmetries respected by the truncated Hamiltonian.
Note that this argument does not mean that the discarded pieces of the Hamiltonian consist of only irrelevant operators. These components will likely contain both relevant and irrelevant operators. As a result, the coupling in the Yang-Mills piece of the Symanzik action will not be given by $g$ from the truncated Hamiltonian. The coupling in the Symanzik action will be renormalized with dependence on the truncation used. This is different from the untruncated case, where the Symanzik action has the same coupling as the Hamiltonian. Connecting simulations done with a truncated Hamiltonian to continuum physics will require a detailed understanding of how the truncation renormalizes the coupling in the Yang-Mills piece.

The lattice spacing $a$ is inversely proportional to the correlation length of the system and goes to zero as the correlation length diverges. 
As~\cref{fig:gluemass} shows, it appears that the correlation length of the truncated Hamiltonians freezes out, which sets a limit on how small a lattice spacing can be reached.
From the perspective of the Symanzik action, the freezing out of the correlation length prevents the coefficients of the irrelevant operators from becoming arbitrarily small, as they do in the continuum limit.
As the only deviations from the continuum limit are suppressed by powers of $a$, simulations with the truncated Hamiltonian can be performed with various values of $a$ up to the minimum value allowed by the truncation, and then extrapolations can be performed to smaller $a$. 
This is the same as how traditional lattice calculations are performed in practice. From this perspective, increasing the truncation of the Hilbert space solely serves the role of making smaller lattice spacings accessible to simulation. 
However, performing these calculations with controlled errors requires a more detailed understanding of the effective actions that describe these truncated Hamiltonians.

\section{Computational Resources}
To simulate the real-time dynamics of this theory, it will be necessary to construct a circuit to implement the time evolution operator. Time evolution will be implemented using a first-order Trotterization  formula, which approximates time evolution by
\begin{equation}
    e^{-i\Delta t \sum_j \hat{H}_j} = \prod_j e^{-i\Delta t  \hat{H}_j} + O (\Delta t^2) \,,
\end{equation}
where each of the $e^{-i\Delta t  \hat{H}_j}$ can be efficiently implemented on quantum hardware. Note that the circuits developed in this section could, in principle, be used to implement higher-order Trotter formulas. In this work, the Hamiltonian will be Trotterized by
\begin{equation}
    e^{-i \hat{H} t} \approx e^{-i \hat{H}_E t} e^{-i \hat{H}_B t} \,.
\end{equation}
All terms in $\hat{H}_E$ commute with each other, so it does not matter in which order they are implemented. 
In the truncated Hamiltonian, plaquette operators with a shared link do not commute, and the order in which their time evolution is implemented matters. 
To implement the plaquette operators, the lattice should be split into separate sublattices where on each sublattice, the plaquettes have no shared links. Each sublattice can then be evolved in parallel.
\begin{table}[]
    \centering
    \begin{tabular}{|c|c|c|c|c|c|c|}
    \hline
             & $(1,1,1)$ $2D$ & $(1,1,1)$ $3D$ & $(1,2,1)$ $2D$ & $(1,2,1)$ $3D$ & $(2,2,2)$ $2D$ at order $1/N_c$ & $(2,2,2)$ $3D$ at order $1/N_c$ \\
    \hline
        Qubits & 1 & 1 & 1 & 1 & 5 & 4 \\ 
    \hline
        CNOT & 16 & 134 & 20 & 1,744 & 249,274 & 249,226 \\ 
    \hline
        H    & 2 & 46 & 2 & 598 & 72,030 & 72,030 \\ 
    \hline
        $R_z$& 17 & 157 & 19 & 2,030 & 247,216 & 247,096 \\ 
    \hline
    \end{tabular}
    \caption{Number of qubits and gates per plaquette required to implement a single first order Trotter step for different truncations of the Hamiltonian. Note that in $2D$ there are two links per plaquette and in $3D$ there is a single link per plaquette.}
    \label{tab:gate_count}
\end{table}

The simplest truncation to simulate in two dimensions is the $(1,1,1)$ truncation in the even $C$ sector. 
The electric terms can be implemented with a single $Z$ rotation per plaquette, and the plaquette terms can be implemented using previously developed circuits~\cite{Ciavarella:2024fzw}. 
The implementation of a single plaquette term will require $16$ CNOT gates, $2$ H gates, and $16$ $R_z$ gates. 
A circuit of the same size can be used to implement the plaquette operator for the $(1,2,1)$ truncation in two dimensions with the qubit encoding. 
An additional $2$ CNOT gates and $R_z$ gate will be needed per link to implement the electric energy operator in the $(1,2,1)$ truncation.

In three spatial dimensions with the $(1,1,1)$ truncation, the electric term can be implemented with a single $R_z$ again, but the plaquette term is complicated by each plaquette on a square 3D lattice having $12$ neighbors instead of $4$ as in 2D. 
This means that implementing the time evolution generated by a single plaquette term requires implementing $\exp(i \theta \hat{P}_0^{\otimes 12} \hat{X}_p )$. 
This can be done using ancilla qubits with textbook techniques~\cite{nielsen2001quantum}. One can use Toffoli gates to encode the state of the control qubits onto ancilla qubits. 
The $X$ rotation on the active qubit will be controlled on a single ancilla qubit. 
The control information on the ancillas will then be uncomputed with Toffoli gates. For a rotation controlled by $12$ qubits, this will require $11$ ancilla qubits and a total of $22$ Toffoli gates. 
The rotation controlled by a single ancilla qubit will require $2$ CNOT gates, $2$ H gates, and $2$ $R_z$ gates. 
With the standard Toffoli circuit, the gate count for a single plaquette operator is $134$ CNOT gates, $46$ H gates, and $156$ $R_z$ gates. 
Note that cancellations can likely be found to reduce this gate count, but in this work, we are interested in only an order-of-magnitude estimate of the resources necessary for performing time evolution. 
With the qubit encoding of the $(1,2,1)$ truncation, the plaquette operator can be implemented using $13$ Givens rotations, which gives the gate count in~\cref{tab:gate_count}.

The next truncation that allows one to approach closer to the continuum limit in two spatial dimensions
is the $(2,2,2)$ truncation. At this truncation, there is a single qubit assigned to each link and a qu$6$ assigned to each plaquette. 
For simulations on qubit hardware, each qu$6$ can be replaced by $3$ qubits. 
The electric term of the Hamiltonian consists of single plaquette terms, and terms involving a plaquette, a link, and a neighboring plaquette. 
The evolution generated by the single plaquette terms can be implemented by decomposing it into a sum over tensor products of Pauli $\hat{Z}$ terms and evolving each one separately. 
This will require $8$ CNOT gates and $8$ $R_z$ gates per plaquette. The electric Hamiltonian terms with support on two plaquettes and a link will take a similar form, except acting on $7$ qubits instead of $3$. 
Using the same strategy, the time evolution generated by these terms can be implemented using $128$ CNOT gates and $128$ $R_z$ gates per link. 
The plaquette term is somewhat more complicated to implement. In this work, the plaquette operator will be decomposed into a sum over generators of gauge-invariant rotations, and each rotation will be implemented independently. 
This approach was previously suggested for performing quantum simulations of SU($3$) lattice gauge theory on a chain of plaquettes~\cite{Ciavarella:2021nmj}. 
Each rotation couples two basis states and can be implemented with standard techniques for Givens rotations. 
These Givens rotations act on $19$ qubits, and a Gray code can be implemented with CNOT gates to turn the Givens rotation into a single qubit rotation controlled by $18$ qubits. 
The Gray code requires at most $36$ CNOT gates to implement and undo. The controlled rotation of the qubit will use $18$ ancilla qubits and $34$ Toffoli gates to encode/decode control information on the ancillas and $2$ H gates, $2$ $R_z$ gates and $2$ CNOT gates to implement the rotation controlled by the ancilla. 
Therefore, the total gate count for a single Givens rotation is $242$ CNOT gates, $70$ H gates, and $240$ $R_z$ gates. 
The number of Givens rotations that need to be performed depends on the order of the large $N_c$ expansion used. \Cref{tab:LargeNTerms} shows the number of Givens rotations required for different truncations in large $N_c$. 
The large $N_c$ expansion reduces the number of Givens rotations needed in the time evolution circuit by roughly $95\%$ from the theory with no large $N_c$ truncation. 
The gate counts for this truncation are shown in~\cref{tab:gate_count}.
\begin{table}[]
    \centering
    \begin{tabular}{|c|c|}
    \hline
         & (2,2,2) Truncation  \\
         \hline
        $1$ & $517$\\
        \hline
        $1/N_c$ & $1029$ \\
        \hline
        No Large $N_c$ truncation & $19,594$ \\
        \hline
    \end{tabular}
    \caption{The number of Givens rotations present when keeping terms in the plaquette order to different powers of $1/N_c$. The number of Givens rotations needed without a large $N_c$ truncation was taken from Ref.~\cite{Ciavarella:2021nmj}.}
    \label{tab:LargeNTerms}
\end{table}

In three spatial dimensions at the $(2,2,2)$ truncation, the electric energy operators can be implemented using the same strategy as in $2$D and the gate counts will be the same per link. 
As with the $(1,1,1)$ truncation, the implementation of plaquette operators is complicated by each plaquette having $3$ neighboring plaquettes per link. 
Due to this truncation, if two of the plaquettes neighboring a single link are excited, there will be no non-zero matrix elements for the plaquette operator. 
Note that the matrix elements of the plaquette operator only depend on the number of neighboring plaquettes excited on each link, not which neighboring plaquette is excited. 
This information can be encoded onto a set of three ancilla qubits using $3$ CNOT gates per neighboring plaquette. 
The Givens rotations needed for implementing the time evolution operator will be controlled by these ancillas and the resulting circuit will be identical to the $2$D case. 
The ancilla qubits will then be uncomputed after the Givens rotations. 
This approach to implementing the time evolution operator while keeping terms up to order $1/N_c$ has the gate cost shown in~\cref{tab:gate_count}. 
Note that it is likely more sophisticated methods of constructing Givens rotations can reduce this gate count further~\cite{Davoudi:2022xmb}.

\begin{table}[]
    \centering
    \resizebox{\columnwidth}{!}{\begin{tabular}{|c|c|c|c|c|c|c|}
    \hline
             & $(1,1,1)$ $10\times10$ & $(1,1,1)$ $10\times10\times10$ & $(1,2,1)$ $10\times10$ & $(1,2,1)$ $10\times10\times10$ & $(2,2,2)$ $10\times10$ at order $1/N_c$ & $(2,2,2)$ $10\times10\times10$ at order $1/N_c$ \\
    \hline
        Qubits & $100$ & $3,000$ & $100$ & $3,000$ & $500$ & $12,000$ \\ 
    \hline
        $T$ & $64,600$ & $1.79\times10^7$ & $72,700$ & $2.3\times10^8$ & $9.4\times10^8$  &  $2.8\times10^{10}$\\ 
    \hline
    \end{tabular}}
    \caption{Number of qubits required to represent the state of the system and $T$ gates required to implement a single first order Trotter step for different truncations of the Hamiltonian on a $10\times10$ and $10\times10\times10$ lattice. For the $T$ gate count, it is assumed that $R_z$ gates are synthesized to an accuracy of $\epsilon=10^{-10}$.}
    \label{tab:Tgate_count}
\end{table}

While the gate count for the $(1,1,1)$ truncation is small enough to be implemented on NISQ devices, the $(2,2,2)$ truncation (and a closer approach to the continuum limit) will likely require using an error-corrected quantum computer. 
This changes the dominant cost of the circuit from CNOT gates to $R_z$ gates. 
In an error-corrected quantum computer, $R_z$ gates are implemented through the use of $T$ gates. The number of $T$ gates required to synthesize an $R_z$ gate to accuracy $\epsilon$ is $1.15\log_2\left(\frac{1}{\epsilon}\right)$~\cite{Chuang:1996hw}. 
Using the results from~\cref{tab:gate_count} and assuming $R_z$ are synthesized to an accuracy of $\epsilon=10^{-10}$, the implementation described in this section require $~9.4\times10^8$ $T$ gates per Trotter step on a $10\times10$ lattice and $~2.8\times10^{10} \ T$ gates per Trotter step on a $10\times10\times10$ lattice with the $(2,2,2)$ truncation with $1/N_c$ corrections. 
Previous work has estimated that physically interesting simulations of the dynamics of lattice QCD could be performed on a $10\times10\times10$ lattice with a spacing of $a\approx 0.1$ fm~\cite{Cohen:2021imf}. 
The two-dimensional simulations performed in the previous section show that the truncations introduced in this work can reach vacuum correlation lengths of roughly two lattice sites. 

In three spatial dimensions, the lattice spacing can be set by matching the correlation length to the predicted scalar glueball mass of $1730$ MeV~\cite{Morningstar:1999rf,Huber:2020ngt,Ochs:2013gi}. 
If similar correlation lengths can be reached in a three-dimensional system, this means the truncations in this work can reach a lattice spacing of $a\approx0.057$ fm. 
The number of $T$ gates required for a simulation was estimated to be $3.01\times10^{49}$ in previous work~\cite{Kan:2021xfc}. Follow-up work using qubitization methods instead of Trotterization reduced this gate count to $1.7\times10^{27}$~\cite{Rhodes:2024zbr}. 
The $(2,2,2)$ truncation reduces the $T$ gate count to $\mathcal{O}(10^{10})$ per Trotter step. 
This lattice spacing may also be accessible to the $(1,2,1)$ truncation, which would have a $T$ gate count of $\mathcal{O}(10^8)$ per Trotter step. The $T$ gate counts for all other truncations are given in~\cref{tab:Tgate_count}. 
Note that these previous estimates include matter as well as the gauge fields; however, the dominant cost of the circuits comes from implementing the gauge fields.
Therefore, it is reasonable to use them for an order-of-magnitude estimate for the gate count. 
Those previous resource estimates were based on rigorous upper bounds for errors in the time evolution circuit~\cite{Childs:2019hts}. 
At long times, these bounds fail to capture the suppression of Trotterization errors due to a form of many-body localization~\cite{Heyl:2019rtj}. 
An estimate of this form will not be done in this work, as numerical studies have demonstrated these bounds are extremely loose at shorter times as well and lead to overestimates of resource requirements by orders of magnitude~\cite{Childs:2018xxl}. 
For example, estimates of this kind suggest $\mathcal{O}(10^{13})$ $T$ gates are necessary to simulate the Schwinger model on $128$ sites~\cite{Sakamoto:2023cxs}, but in reality, $112$ site simulations have been successfully performed on NISQ hardware~\cite{Farrell:2023fgd,Farrell:2024fit}. 
Additionally, by treating the Trotterized time evolution operator as a temporal lattice, one can approach the $\Delta t \rightarrow 0$ limit as the spatial lattice spacing is taken to $0$. 
This should be a more economical approach to simulation than approaching $\Delta t \rightarrow 0$ for each lattice spacing as proposed in Ref.~\cite{Carena:2021ltu}. 

Recent work has also estimated the number of $T$ gates needed to simulate the harshest truncation of SU($3$) lattice gauge theory in the electric basis without a large $N_c$ expansion~\cite{Balaji:2025afl}. The $(1,1,1)$ and $(1,2,1)$ truncations introduced in this work should reproduce physics with an accuracy similar to this truncation. With the use of a large $N_c$ truncation, the $T$ gate cost of a Trotter step on a $10\times10$ lattice in the $(1,2,1)$ truncation is an order of magnitude smaller than that for a $3\times3$ lattice with no large $N_c$ expansion~\cite{Balaji:2025afl}. In three spatial dimensions, the $T$ gate cost of a Trotter step in the $(1,1,1)$ truncation on a $10\times10\times10$ lattice is comparable to the cost of a Trotter step with no large $N_c$ expansion on a $3\times3\times3$ lattice. Note that the techniques used to optimize circuits in Ref.~\cite{Balaji:2025afl} could be used to improve the gate counts for the circuits in this work.

\section{Discussion}

The large $N_c$ expansion introduced in Ref.~\cite{Ciavarella:2024fzw} has been extended to include $1/N_c$ corrections. 
At this truncation in large $N_c$, states can have extended loops of flux as well as loops around single plaquettes. 
The Hilbert space was truncated using a local Krylov basis construction. The form of the lowest lying truncations were given. 
The scalar glueball mass was computed in these truncated Hamiltonians with tensor networks which showed that these truncations may be able to reach smaller lattice spacings than one would initially expect. 
In traditional lattice calculations, the use of smeared actions (such as those generated by Wilson flows) reduces the size of lattice artifacts~\cite{Luscher:2010iy}. 
In the Hamiltonian formulation, the analogous procedure can be performed using the similarity renormalization group, and applying it to these truncations should enable reaching even smaller lattice spacings~\cite{Ciavarella:2023mfc}. 
Explicit strategies for constructing the time evolution operator on a quantum computer were given for the low-lying truncations in $2+1D$ and $3+1D$. 
The resource requirements for performing time evolution are orders of magnitude smaller than previous approaches that used a direct encoding of the gauge fields onto qubits. The circuit requirements for moderately sized $2+1$D systems are within the $5$ year roadmaps of multiple quantum computing companies~\cite{IBMRoadmap,quantinuumRoadmap}.
Note that it is likely possible to reduce requirements further with more efficient circuit constructions. 
It may also be possible to use native gates of specific quantum computing platforms to reduce resources further. 
Future work will also explore how to use the $1/N_c$ suppression of some terms in the Hamiltonian to more efficiently implement the time evolution operator. 
By reducing the resources required for simulation, this work brings quantum simulations relevant to high energy physics closer to reality. 
To reach the goal of simulating full QCD on a quantum computer, it will be necessary to include fermionic matter in this formulation. 
It is expected that the simplifications obtained in this work will carry over to the inclusion of matter.

\begin{acknowledgments}
The authors would like to acknowledge many useful discussions with Jesse Stryker, Irian D'Andrea, and Neel Modi. This material is based upon work supported by the U.S. Department of Energy, Office of Science, National Quantum Information Science Research Centers, Quantum Systems Accelerator. Additional support is acknowledged from the U.S. Department of Energy (DOE), Office of Science under contract DE-AC02-05CH11231, partially through Quantum Information Science Enabled Discovery (QuantISED) for High Energy Physics (KA2401032). This research used resources of the National Energy Research Scientific Computing Center, which is supported by the Office of Science of the U.S. Department of Energy under Contract No. DE-AC02-05CH11231. All data displayed in figures is available as supplemental material in the arXiv submission.
\end{acknowledgments}

\appendix

\section{Vanishing Correlations in the (1,2,2) Truncation}
\label{app:12Decouple}
In the main text, it was shown that the $(1,2,2)$ truncation of the Hamiltonian completely fails to reproduce the glueball mass. As explained in this section, this is due to this truncation being close to a point in theory space where the correlation length is $0$ for all values of $g$. If one considers the $(1,2,2)$ truncation with the $1/N_c$ splitting in the Casimirs dropped and the plaquette term in the $N_c\rightarrow\infty$ limit, each plaquette decouples from the others.

To determine the form of the Hamiltonian (for the even $C$ sector) truncated in this manner, we will first consider a lattice composed of a single plaquette. For a lattice of this size, gauge-invariant states can be labeled by the representation $\mathbf{R}$ flowing around the plaquette. Using the known expressions for matrix elements of the plaquette operator, we find
\begin{align}
    \left(\hat{\Box} + \hat{\Box}^\dagger \right) \ket{\mathbf{1}} & = \ket{\wh} + \ket{\bl} \,.
\end{align}
We can represent the state of a single plaquette with a qubit with basis states.
\begin{align}
    \ket{0} & = \ket{\mathbf{1}} \nonumber \\ 
    \ket{1} & = \frac{1}{\sqrt{2}} \left(\ket{\wh} + \ket{\bl} \right) \,.
\end{align}
With these basis states, the Hamiltonian is
\begin{equation}
    \hat{H} = N_c g^2 \hat{P}_1 - \frac{1}{g^2\sqrt{2}}  \hat{X} \,,
\end{equation}
where $\hat{P}_i = \ket{i}\bra{i}$.
%At leading order in $1/N_c$, the Casimirs of the representations $\blwh$, $\whwhs$ and $\whwha$ are the same so 

For a two plaquette system with open boundary conditions (and truncating as described above), gauge invariant states can be specified using the representation on the three vertical links, $\ket{\mathbf{R}_1,\mathbf{R}_2,\mathbf{R}_3}$. The physical states that need to be represented at this truncation can once again be determined by applying plaquette operators to the electric vacuum state $\ket{\mathbf{1},\mathbf{1},\mathbf{1}}$. A single application of either of the plaquette operators can create the states
\begin{align}
    \left(\hat{\Box}_1 + \hat{\Box}_1^\dagger \right) \ket{\mathbf{1},\mathbf{1},\mathbf{1}} &= \ket{\wh,\bl,\mathbf{1}} + \ket{\bl,\wh,\mathbf{1}} \nonumber \\
    \left(\hat{\Box}_2 + \hat{\Box}_2^\dagger \right) \ket{\mathbf{1},\mathbf{1},\mathbf{1}} &= \ket{\mathbf{1},\wh,\wh} + \ket{\mathbf{1},\bl,\bl} \,.
\end{align}
Applying neighboring plaquette operators produces the state
\begin{align}
   \left(\hat{\Box}_2 + \hat{\Box}_2^\dagger \right) \left(\hat{\Box}_1 + \hat{\Box}_1^\dagger \right) \ket{\mathbf{1},\mathbf{1},\mathbf{1}} = & \left(\hat{\Box}_2 + \hat{\Box}_2^\dagger \right) \left(\ket{\wh,\bl,\mathbf{1}} + \ket{\bl,\wh,\mathbf{1}} \right) \nonumber \\
   = & \sqrt{1 - \frac{1}{N_c^2}} \ket{\wh,\blwh,\wh} + \frac{1}{N_c} \ket{\wh,\mathbf{1},\wh} + \sqrt{1 - \frac{1}{N_c^2}} \ket{\bl,\blwh,\bl} + \frac{1}{N_c} \ket{\bl,\mathbf{1},\bl}  \nonumber \\
   & + \sqrt{\frac{1}{2} \left( 1 + \frac{1}{N_c}\right)} \ket{\bl,\whwhs,\wh} + \sqrt{\frac{1}{2} \left( 1 - \frac{1}{N_c}\right)} \ket{\bl,\whwha,\wh} \nonumber \\
   & + \sqrt{\frac{1}{2} \left( 1 + \frac{1}{N_c}\right)} \ket{\wh,\blbls,\bl} + \sqrt{\frac{1}{2} \left( 1 - \frac{1}{N_c}\right)} \ket{\wh,\blbla,\bl} \,.
\end{align}
In the large $N_c$ limit this state becomes
\begin{align}
   \left(\hat{\Box}_2 + \hat{\Box}_2^\dagger \right) \left(\hat{\Box}_1 + \hat{\Box}_1^\dagger \right) \ket{\mathbf{1},\mathbf{1},\mathbf{1}}  = & 
  \ket{\wh,\blwh,\wh}  +  \ket{\bl,\blwh,\bl}   + \frac{1}{\sqrt{2}}\ket{\bl,\whwhs,\wh} + \frac{1}{\sqrt{2}} \ket{\bl,\whwha,\wh}  \nonumber \\ &+ \frac{1}{\sqrt{2}} \ket{\wh,\blbls,\bl} +\frac{1}{\sqrt{2}} \ket{\wh,\blbla,\bl} \,.
\end{align}
At this truncation, there are no additional states for a lattice of this size. Gauge-invariant states of this lattice can be specified using a qubit per plaquette. When one qubit is in the $0$ state, the state assignment for the other qubit is identical to the one-plaquette cases. At leading order in $1/N_c$, the Casimirs of the representations $\blwh$, $\whwhs$ and $\whwha$ are the same so it is not necessary to specify which of these irreps is on the vertical link. The state $\ket{1,1}$ corresponds to
\begin{equation}
    \ket{1,1} = \frac{1}{2} \left(\ket{\wh,\blwh,\wh}  +  \ket{\bl,\blwh,\bl}   + \frac{1}{\sqrt{2}}\ket{\bl,\whwhs,\wh} + \frac{1}{\sqrt{2}} \ket{\bl,\whwha,\wh}  + \frac{1}{\sqrt{2}} \ket{\wh,\blbls,\bl} +\frac{1}{\sqrt{2}} \ket{\wh,\blbla,\bl} \right) \,.
\end{equation}
This basis construction scales to larger 2D lattices, using a single qubit per plaquette. With this basis, the Hamiltonian is given by
\begin{equation}
    \hat{H}_2 = \sum_p \left(N_c g^2 \hat{P}_{1}^p  - \frac{1}{g^2\sqrt{2}}  \hat{X}^{p}\right) \,.
\end{equation}
From this expression, it is clear that each plaquette in the lattice is independent of all of the others, and the system is completely uncorrelated. 

The $(1,2,2)$ Hamiltonian differs from this by terms in the electric Hamiltonian that are $\mathcal{O}\left(\frac{g^2}{N_c}\right)$ and terms in the magnetic Hamiltonian that are $\mathcal{O}\left(\frac{1}{g^2 N_c}\right)$. 
The energy gap between eigenstates of $\hat{H}_2$ is given by $2\sqrt{N_c^2 g^4/4 + 1/(2g^4)}$ so $\hat{H}^{(1,2,2)}$ can be treated as being perturbatively close to $\hat{H}_2$ for any value of $g$. Note that this argument only applies to the (1,2,2) truncation, as the other truncations will have additional terms in the Hamiltonian that are comparable in size to the gap of $\hat{H}_2$. As the following argument will show, being perturbatively close to $\hat{H}_2$ is sufficient to guarantee that the correlation length in the $(1,2,2)$ truncation stays small at all values of $g$.

For a generic lattice Hamiltonian 
\begin{equation}
    \hat{H} = \hat{H}_0 + \varepsilon \hat{V} \,,
\end{equation}
where $\hat{H}_0$ only acts on single sites, $\hat{V}$ couples nearest neighbor sites, and $ \varepsilon$ is a small number, it can be shown in perturbation theory that the leading contribution to a connected vacuum correlation function is 
\begin{equation}
    G(x) = \bra{\psi} \mathcal{O}(x) \mathcal{O}(0) \ket{\psi} -\bra{\psi} \mathcal{O}(0) \ket{\psi}^2 = \varepsilon^{\abs{x}} + \mathcal{O}(\varepsilon^{\abs{x}+1}) \,.
\end{equation}
This follows from applying traditional perturbation theory to the eigenstates of $\hat{H}_0$. Any term below order $\abs{x}$ will only contribute to the connected piece of the correlation function. As a result, the effective mass obtained from such a correlation function is $m_{eff} = \log\left(\frac{1}{\varepsilon}\right)$.

From this result, it can be argued that for any value of $g$, the glueball mass for the $(1,2,2)$ truncation will be large. This is an indication that this is not a useful truncation for attempting to approach the continuum limit, as the lattice spacing will be set by the inverse of the glueball mass. If the glueball mass does not approach zero (as is the case in this truncation), then the continuum limit cannot be approached.

\section{Strictly leading order in \texorpdfstring{$1/N_c$}{1/Nc}: An exactly solvable model}
\label{app:leading_Nc}
We now turn to the decoupling of plaquettes mentioned around \eqref{eq:trivial_hilbert}.
This is obtained when only the leading $N_c$ dependence of the electric energy and plaquette operators is kept, and is a general feature of the Kogut-Susskind Hamiltonian, unrelated to truncation schemes of the electric basis.
We flag to the reader that this is a different phenomenon than the decoupling studied in \cref{app:12Decouple}.
For a single plaquette, one can define electric basis states $\ket{n}$ by a sum over all representations whose Young tableaux have $n$ boxes.
Starting from the electric vacuum, these are the only states that can be obtained through time evolution at this order.
Different representations will have coefficients proportional to the number of ways they can be built from lower-weight representations. Explicitly, the lowest lying basis states are
\begin{align}
    \ket{0} &= \ket{\mathbf{1}} \nonumber \\
    \ket{1} &= \frac{1}{\sqrt{2}} \left(\ket{\wh} + \ket{\bl}\right) \nonumber \\
    \ket{2} &= \frac{1}{\sqrt{8}} \left(\ket{\whwhs} + \ket{\whwha} + \ket{\blbls}+\ket{\blbla}+2\ket{\blwh}\right) \ \ \ .
\end{align}
With this basis, one can show that the Hamiltonian for a single plaquette is
\begin{equation}
    \hat{H} = g^2 N_c\sum_n n \hat{P}_n - \frac{1}{\sqrt{2}g^2} \sum_n \sqrt{n}\, \hat{\mathcal{X}}_{n,n+1} \,,
\end{equation}
where $\hat{P}_n=\ket{n}\bra{n}$ and $\hat{\mathcal{X}}_{n,n+1} = \ket{n}\bra{n+1} + \ket{n+1}\bra{n}$. This Hamiltonian is a displaced harmonic oscillator. This can be seen by introducing the usual raising and lowering operators, and corresponding $\hat x$ and $\hat p$ operators. In particular,
\begin{align}
    \sum_n n \hat P_n &= a^\dagger a\nonumber\\
    \sum_n \sqrt{n} \hat{\mathcal{X}}_{n,n+1} &= a + a^\dagger
    \,,
\end{align}
which allows the Hamiltonian to be written as
\begin{align}
    \hat H = \lambda \left(a^\dagger a\right)- \frac{1}{\sqrt{2}g^2} (a + a^\dagger)\,,
    \label{eq:Hamiltonian1}
\end{align}
with 
\begin{align}
    \lambda \equiv g^2 N_c
    \,.
\end{align}
Defining
\begin{align}
    \hat x = \sqrt{\frac{1}{2 \lambda}}(a + a^\dagger)\,, \qquad 
    \hat p = i \sqrt{\frac{\lambda}{2}}(a^\dagger - a)
    \,,
\end{align}
we can write
\begin{align}
    \hat H &= \frac{\hat p^2}{2} + \frac{ \lambda^2}{2} \hat x^2 - \frac{\sqrt{ \lambda}}{g^2}  \hat x - \frac{\lambda}{2} \nonumber\\
    \hat H &= \frac{\hat p^2}{2} + \frac{ \lambda^2}{2} \left(\hat x - \frac{\sqrt{ \lambda}}{g^2 \lambda^2} \right)^2 - \frac{\lambda}{2} - \frac{1}{2g^4 \lambda} \nonumber\\
    \label{eq:Hamiltonian}
\end{align}
The spectrum of this Hamiltonian can be obtained immediately, since this is just a harmonic oscillator with a shifted ground state. 
One can read off the eigenvalues from Eq.~\eqref{eq:Hamiltonian}
\begin{align}
    E_n = \lambda  n - \frac{1}{2 g^4 \lambda}\,,
\end{align}

The corresponding eigenstates are obtained from the states $\ket{n}$ by using the displacement operator
\begin{align}
    D(\alpha) = \exp\left[\alpha(a^\dagger - a)\right]
    \,,
\end{align}
which satisfies
\begin{align}
    a D(\alpha) = D(\alpha) (a + \alpha)\,, \qquad a^\dagger D(\alpha) = D(\alpha) (a^\dagger + \alpha)
    \,.
\end{align}
Defining 
\begin{align}
    (a^\dagger - \alpha)(a-\alpha) \ket{n, \alpha} 
    &= a^\dagger a - \alpha(a^\dagger + a) - \alpha^2 \nonumber\\
    &= (n - \alpha^2)\ket{n, \alpha}
    \,,
\end{align}
one can show that 
\begin{align}
    a^\dagger a - \alpha(a^\dagger + a) \ket{n, \alpha}
    &= \left[(a^\dagger - \alpha)(a-\alpha) - \alpha^2\right] \ket{n, \alpha}\nonumber\\
    &= D(\alpha)a^\dagger D^\dagger(\alpha)D(\alpha)a D^\dagger D(\alpha) \ket{n} - \alpha^2 \ket{n, \alpha}\nonumber\\
    &= D(\alpha)a^\dagger a \ket{n} - \alpha^2 \ket{n, \alpha}\nonumber\\
    &= (n - \alpha^2)\ket{n, \alpha}
    \,.
\end{align}
This implies that the Hamiltonian in~\eqref{eq:Hamiltonian1} is diagonalized by the displaced states
\begin{align}
    \ket{n, \lambda, g} \equiv D\left(-\frac{1}{\sqrt{2\lambda} g^2}\right) \ket{n}  \,.
\end{align}
On a larger lattice, it can be shown that to leading order in large $N_c$, plaquette matrix elements are independent of the state of the neighboring plaquettes~\cite{Ciavarella:2024fzw}. Using this result, it follows that in this limit, the Kogut-Susskind Hamiltonian is given by a displaced harmonic oscillator at each plaquette. These harmonic oscillators are not coupled, leading to a system with zero correlation length.

\section{Dimension and Casimir of general SU\texorpdfstring{$(N)$}{(N)} representations}
\label{app:casimirs}
Using the results given in the appendix of~\cite{Sannino2013irreps}, the dimension and Casimir of a general SU($N$) irreducible representation with Dynkin indices $(m_1, \ldots , m_{N-1})$ can be written in the compact notation
\begin{align}
    d(m_1, \ldots , m_{N-1}) &= \prod_{p=1}^{N-1} \left[ \frac{1}{p!} \prod_{q=p}^{N-1} \left( \sum_{z=q-p+1}^p (1+m_z)\right)\right]\nonumber\\
    C(m_1, \ldots , m_{N-1})&= \frac{1}{2N} \sum_{k=1}^{N-1} \left[ N k(N-k) m_k + k(N-k) m_k^2 + \sum_{l=0}^{k-1} 2l(N-k) m_l m_k \right]
    \,.
\end{align}
Using these expressions, one can immediately obtain the results for the representations used in this work, which we present in tabulated form, both for general $N$ and for $N=3$

\begin{table}[h!]\begin{tabular}{|c|c|c|c|c|}
\hline
Irrep & \multicolumn{2}{c|}{Dimension} & \multicolumn{2}{c|}{Casimir}\\
 & General $N$ & $N = 3$ & General $N$ & $N = 3$ \\\hline
 $(0,0,  \ldots, 0, 0)$ & $1$& $1$& $0$ & $0$ \\
 $(1,0,  \ldots, 0, 0)$ & $N$& $3$& $(N^2-1)/(2N)$& $4/3$\\
 $(0,0,  \ldots, 0, 1)$ & $N$& $3$& $(N^2-1)/(2N)$& $4/3$\\
 $(2,0,  \ldots, 0, 0)$ & $N(N+1)/2$& $6$& $(N^2+N-2)/N$&$10/3$\\
 $(0,0,  \ldots, 0, 2)$ & $N(N+1)/2$& $6$& $(N^2+N-2)/N$&$10/3$\\
 $(0,1,0  \ldots, 0)$ & $N(N-1)/2$& 3 & $(N^2-N-2)/N$&$4/3$\\
 $(0,  \ldots, 0,1, 0)$ & $N(N-1)/2$& $3$& $(N^2-N-2)/N$ &$4/3$\\
 $(1,0,  \ldots, 0, 1)$ & $N^2-1$& $8$& N & 3\\\hline
\end{tabular}
\end{table}

\section{Plaquette Matrix Elements}
\label{app:clebsch_gordan}
\begin{figure}
    \centering
    \begin{tikzpicture}
        \filldraw (0,0) circle (3pt);
        \filldraw (0,2) circle (3pt);
        \filldraw (2,0) circle (3pt);
        \filldraw (2,2) circle (3pt);

        \draw (0,0) -- node[anchor=east] {$R_1$} ++(0,2);
        \draw (0,2) -- node[above] {$R_2$} ++(2,0);
        \draw (2,0) -- node[anchor=west] {$R_3$} ++(0,2);
        \draw (2,0) -- node[below] {$R_4$} ++(-2,0);

        \draw (-0.5,-0.5) node[anchor=east]{$C_4$} -- (0,0);
        \draw (2.5,-0.5) node[anchor=west]{$C_3$} -- (2,0);
        \draw (-0.5,2.5) node[anchor=east]{$C_1$} -- (0,2);
        \draw (2.5,2.5) node[anchor=west]{$C_2$} -- (2,2);
    \end{tikzpicture}
    \caption{A single plaquette with one external link at each vertex.}
    \label{fig:plaquette}
\end{figure}
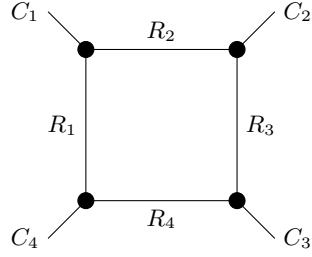

The explicit form of the SU($N_c$) Hamiltonians in this work requires the computation of matrix elements of the plaquette operator between gauge-invariant states. In previous work, it was noted that the state of a plaquette with a single external link at each vertex (as shown in~\cref{fig:plaquette}) can be specified by basis states where the irreps on each link are given~\cite{Ciavarella:2021nmj,Ciavarella:2024fzw}, $    \ket{\psi \begin{pmatrix}
        C_1 & R_2 & C_2 \\
        R_1 &     & R_3 \\
        C_4 & R_4 & C_3
    \end{pmatrix}}$ .
A limited number of plaquette matrix elements were also computed in the supplemental material of Ref.~\cite{Ciavarella:2024fzw},
\begin{equation}
    \bra{\psi \begin{pmatrix}
        \mathbf{\Bar{A}} & \mathbf{1} & \mathbf{1} \\
        \mathbf{A} &     & \mathbf{1} \\
        \mathbf{A} & \mathbf{1} & \mathbf{1}
    \end{pmatrix}} \hat{\Box} \ket{\psi \begin{pmatrix}
        \mathbf{\Bar{A}} & \mathbf{R} & \mathbf{1} \\
        \mathbf{B} &     & \mathbf{\Bar{R}} \\
       \mathbf{A} & \mathbf{\Bar{R}} & \mathbf{1}
    \end{pmatrix}} = \sqrt{\frac{\text{dim}(\mathbf{B})}{\text{dim}(\mathbf{A}) \text{dim}(\mathbf{R})}}\,,    \label{eq:plaq_elements}
\end{equation}
where the plaquette operator has parallel transporters in representation $\mathbf{R}$ and $\mathbf{B}\in \mathbf{R} \otimes \mathbf{A}$~\cite{Ciavarella:2024fzw}. Other plaquette matrix elements were computed by applying neighboring plaquette operators to create different initial states and using the fact that neighboring plaquette operators commute. The same technique can be used to determine the plaquette matrix elements used in the low-lying truncations in this work. At the $(1,2,2)$ truncation, at most one plaquette operator will be applied to each neighboring plaquette. Combined with the fact that plaquette operators commute, it can be seen that the matrix elements at this truncation are given by products of the expression in Eq.~\eqref{eq:plaq_elements}. This is the same method that was used to compute plaquette matrix elements in the supplemental material of Ref~\cite{Ciavarella:2024fzw}.

\section{DMRG Details}
\label{app:DMRG}
\begin{figure}
    \centering
    \includegraphics[width=0.5\linewidth]{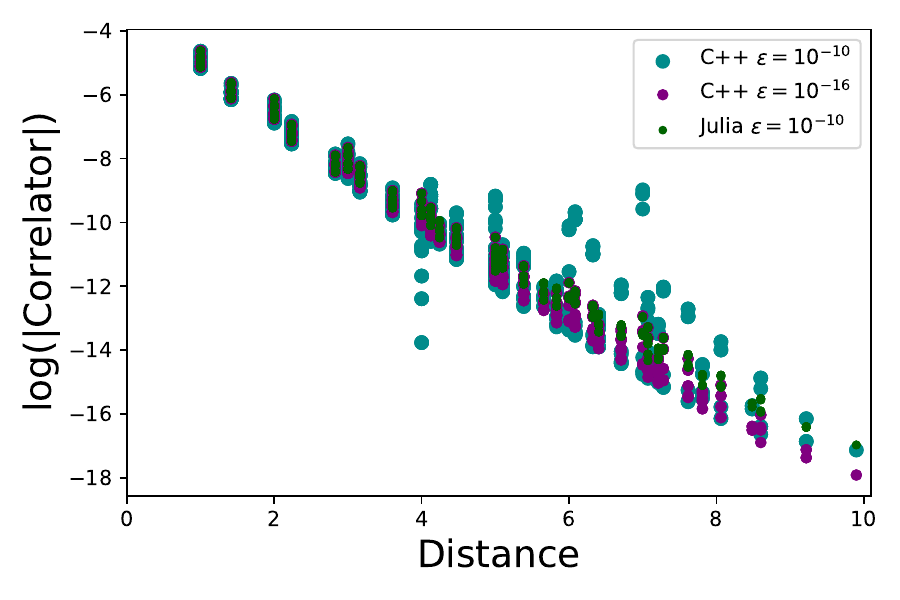}
    \caption{Correlation function for the $(1,1,1)$ truncation on an $8\times8$ lattice with $g=0.8$ computed in C++ and Julia.}
    \label{fig:CorrComparison}
\end{figure}

The DMRG calculations in this work were performed by representing the state as a matrix product state (MPS). The plaquette at coordinates $(x,y)$ on an $L\times L$ lattice corresponds to the tensor at position $y+L(x-1)$, i.e., the MPS winds through the lattice like a typewriter. For the $(1,1,1)$ truncation, $60$ sweeps of the DMRG were performed and the maximum allowed bond dimension was $300$ with a precision cutoff of $10^{-16}$. For the $(1,2,1)$ truncation, $40$ sweeps of the DMRG were performed and the maximum allowed bond dimension was $250$ with a precision cutoff of $10^{-16}$. In both the $(1,1,1)$ and $(1,2,1)$ calculations, the bond dimension did not reach the maximum allowed value for the couplings used. For the higher truncations, the link qubits left of and below a plaquette were blocked with the plaquette qudit into a single site in the MPS. DMRG for the higher truncations was performed using $25$ sweeps and a maximum bond dimension of $100$. The $(1,2,2)$ calculations did not ever reach this maximum bond dimension. A typical truncation error for the $(2,2,2)$ simulations at this maximum bond dimension was $~10^{-8}$. The DMRG was initialized with the system in the electric vacuum state.
\begin{figure}
    \centering
    \includegraphics[width=0.5\linewidth]{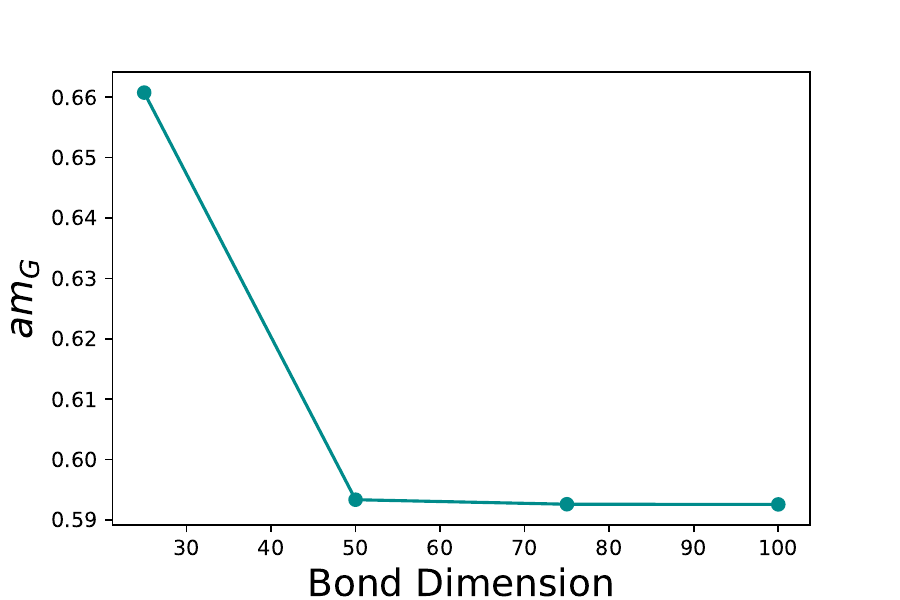}
    \caption{Glueball mass computed in the $(2,2,2)$ truncation on a $4\times4$ lattice with $g=0.2$ as a function of maximum bond dimension.}
    \label{fig:GlueBond}
\end{figure}
The number of sweeps of the DMRG was chosen such that the ground state energy had converged to at least $3$ decimal places for the smallest value of $g$ used. To ensure that a sufficient bond dimension was used for the $(2,2,2)$ truncation, the DMRG was run with multiple different maximum bond dimensions. The resulting glueball mass for a $4\times4$ lattice with $g=0.2$ is shown in~\Cref{fig:GlueBond}. As this figure shows, the glueball mass has converged to sufficient accuracy with a maximum bond dimension of $100$. Similar behavior was found for other values of $g$. For the $8\times8$ calculations, discrepancies in the correlation function were found between the Julia and C\texttt{++} implementations when the same cutoff precision was used. \Cref{fig:CorrComparison} shows that for the $(1,1,1)$ truncation on an $8\times8$ lattice with $g=0.8$, the correlation function computed in C\texttt{++} and Julia with a cutoff precision of $\epsilon=10^{-10}$ differ for distances larger than $4$. When the C\texttt{++} cutoff is decreased to $\epsilon=10^{-16}$, the size of these disagreements decreases, indicating that this is likely a numerical stability issue in the C\texttt{++} implementation. 

\bibliography{ref}

\end{document}